\documentclass{elsarticle}
\usepackage{lmodern}
\usepackage{amssymb}
\usepackage{multicol}
\usepackage{xcolor}
\usepackage{float}
\usepackage[titletoc]{appendix}
\usepackage{amsmath}
\usepackage{graphics}
\usepackage[T1]{fontenc}
\usepackage{rotate}
\usepackage{color}
\usepackage{wrapfig}
\usepackage{latexsym,epsfig,fancyhdr}
\usepackage{latexsym,epsfig}
\graphicspath{{Figures/}}
\usepackage[english]{babel}

\usepackage[margin=2.5cm]{geometry}

\usepackage[bottom]{footmisc}% places footnotes at page bottom
\usepackage{amsmath, amsfonts, amssymb}
\usepackage{tikz}
\usepackage{pgfplots}
\usepackage{pgfplotstable}
\usepackage{booktabs}
\usepackage{multirow}
\usepackage{lscape}
\usepackage{placeins}
\usepackage{mathrsfs}
\usepackage{ulem}
\usetikzlibrary{arrows}

\def\b_v{\mathbf{v}}

\def\V{\mathbf{V}}

\usepackage{pgfplotstable}
\pgfplotsset{compat=1.16}
\usepackage{microtype}
\usepackage{empheq,etoolbox}
\def\og{\leavevmode\raise.3ex\hbox{$\scriptscriptstyle\langle\!\langle$~}}
\def\fg{\leavevmode\raise.3ex\hbox{~$\!\scriptscriptstyle\,\rangle\!\rangle$}}

\begin{document}

\begin{frontmatter}

\selectlanguage{english}
%\title{Assessing the accuracy of the ALE method implemented in the engineering code TrioCFD in the parametric studies of the added mass and damping coefficient induced by a vibrating cylinder in a viscous fluid}
\title{FSI - Vibrations of immersed cylinders. Simulations with the engineering open-source code TrioCFD. Test cases and experimental comparisons}

\selectlanguage{english}
\author[Affil1]{Domenico Panunzio}
\address[Affil1]{Den-Service d'Etudes M\'ecaniques et Thermiques (SEMT), CEA, Universit\'e Paris-Saclay, F-91191, Gif-sur-Yvette, France}
%\address[Affil1]{École Nationale Supérieure d'Arts et Métiers ParisTech (ENSAM), 75013, Paris, France}

\author[Affil2]{Maria Adela Puscas}
\address[Affil2]{Den-Service de Thermohydraulique et de M\'ecanique des Fluides (STMF), CEA, Universit\'e Paris-Saclay, F-91191, Gif-sur-Yvette, France}

\author[Affil1]{Romain Lagrange}
\ead{Corresponding author: romain.lagrange@cea.fr}
%\ead{romain.g.lagrange@gmail.com}

% if you know the dates of reception, and acceptation you can put them now;
% idem for the name of the person presenting the Note

%\medskip
%\begin{center}
%{\small Received *****; accepted after revision +++++\\
%Presented by *****}
%\end{center}

\begin{abstract}
    In this paper, we assess the capabilities of the Arbitrary Lagrangian-Eulerian method implemented in the open-source code TrioCFD to tackle down two fluid-structure interaction problems involving moving boundaries.  
    To test the code, we first consider the bi-dimensional case of two coaxial cylinders moving in a viscous fluid. We show that the two fluid forces acting on the cylinders are in phase opposition, with amplitude and phase that only depend on the Stokes number, the dimensionless separation distance and the Keulegan-Carpenter number. Throughout a detailed parametric study, we show that the self (resp. cross) added mass and damping coefficients decrease (resp. increase) with the Stokes number and the separation distance. Our numerical results are in perfect agreement with the theoretical predictions of the literature, thereby validating the robustness of the ALE method implemented in TrioCFD.
    Then, we challenge the code by considering the case of a vibrating cylinder located in the central position of a square tube bundle. In parallel to the numerical investigations, we also present a new experimental setup for the measurement of the added coefficient, using the direct method introduced by Tanaka. The numerical predictions for the self-added coefficients are shown to be in very good agreement with a theoretical estimation used as a reference by engineers. A good agreement with the experimental results is also obtained for moderate and large Stokes numbers, whereas an important deviation due to parasitic frequencies in the experimental setup appears for low Stokes number. 
    Still, this study clearly confirms that the ALE method implemented in TrioCFD is particularly efficient in solving fluid-structure interaction problems. As an open-source code, and given its ease of use and its flexibility, we believe that TrioCFD is thus perfectly adapted to engineers who need simple numerical tools to tackle down complex industrial problems. 
\end{abstract}

%{\it To cite this article: R. Lagrange, ? (2016).}

%\vskip 0.5\baselineskip

\begin{keyword}
    {Vibration; Fluid-structure interaction; Added mass; Added damping; Stokes number; ALE method; TrioCFD}
\end{keyword}

\end{frontmatter}

\selectlanguage{english}

\section{Introduction}
The accurate estimation of the force acting on a body moving in a viscous fluid is of crucial importance for engineers, in particular for those working in fields such as turbomachinery~\cite{Furber1979}, heat exchangers tube banks~\cite{Chen1975b,Chen1977} or energy harvesting of flexible structures~\cite{Doare2011,Singh2012,Michelin2013,Virot2016,Eloy2008}. When the immersed body is subjected to a small amplitude of motion, Stokes~\cite{Stokes1850} showed that the fluid force is the sum of two terms: an added mass term related to the body's acceleration and a damping term related to the body's velocity. The concept of added mass and damping terms can be generalized to multiple immersed bodies, introducing some self and cross-added coefficients. In such a case, the self-added coefficients relate the fluid force on a body to its motion. On their hand, the cross-added coefficients relate the forces on a body to the other bodies' motion. The determination of the added coefficients has been the topic of considerable experimental,~\cite{1,3,4,6,12,13,17,20,29} and theoretical studies, mostly based on a Helmholtz decomposition associated with a method of images~\cite{Carpenter1958,Landweber1991,Hicks1879,Greenhill1882,Basset1888,
Birkhoff1960,Gibert1980} or a conformal transformation~\cite{Wang2004,Burton2004,Tchieu2010,Scolan2008,Crowdy2006,Crowdy2010,Lagrange2018,Lagrange2020}.

\begin{nomenclature}
\begin{deflist}[AAAAAAAAA] %[AAAA] if you have 4 letters max for example  
\defitem{$\mathbf{v}_{ALE}$}\defterm{Velocity of the ALE frame reference}
\defitem{$J$}\defterm{Jacobian of the transformation between the ALE and the Lagrange frame reference}
\defitem{$\Delta t$}\defterm{Time step of the numerical simulation}
\defitem{$\Delta t_{stab}$}\defterm{Time step which ensure the stability of the numerical simulation at each iteration}
\defitem{$\Delta t_{conv}$}\defterm{Convection time step limit of the numerical simulation}
\defitem{$\Delta t_{diff}$}\defterm{Diffusion time step limit of the numerical simulation}
\defitem{$\Delta t_{max}$}\defterm{Maximum time step of the numerical simulations set by the user}
\defitem{$CFL$}\defterm{Courant-Friedrichs-Lewy number}
\defitem{$\mathbf{v}_h, p_h$}\defterm{Discrete fluid flow velocity vector and pressure}
\defitem{$\mathbf{V}_h, \mathbf{P}_h$}\defterm{Unknowns vectors of the discrete fluid flow velocity vector and pressure}
\defitem{$\mathcal{T}_h$}\defterm{Triangulation of the fluid domain}
\defitem{${K}_j$}\defterm{Triangular element of the discretized fluid domain}
\defitem{$\mathbf{x}_{i}$}\defterm{Middle points of the edges of the triangles}
\defitem{${w}_i$}\defterm{Control volume associated to $\mathbf{x}_{i}$}
\defitem{$\chi_{K_j}, \phi_i$}\defterm{Indicator functions of the triangle $K_j$ and the control volume $w_i$}
\defitem{$[M]$}\defterm{Mass matrix operator}
\defitem{$[A]$}\defterm{Discrete diffusion operator}
\defitem{$[L(\V_h)]$}\defterm{Non-linear discrete convection operator}
\defitem{$[G]$}\defterm{Discrete gradient operator}
\defitem{$[D]$}\defterm{Discrete divergence operator}
\defitem{$lc, lc_{fine}$}\defterm{Largest and smallest mesh size of the fluid domain}
\defitem{$\Delta$}\defterm{Maximum ratio between the lengths of two adjacent edges}
\defitem{$D_j$}\defterm{Diameter of cylinder ${{C}_j}$}
\defitem{$\partial C_j$}\defterm{Boundary of cylinder ${{C}_j}$}
\defitem{$P$}\defterm{Pitch between the cylinders}
\defitem{$\Omega$}\defterm{Angular frequency of the cylinders} 
\defitem{$t, t^*$}\defterm{Dimensional and dimensionless time} 
\defitem{$\partial {C_j}$}\defterm{Boundary of cylinder ${{C}_j}$} 
\defitem{${{\bf{n}}_j}$}\defterm{Outward normal unit vector to cylinder $\partial {C_j}$} 
\defitem{$\rho$}\defterm{Fluid mass density}
\defitem{$\nu$}\defterm{Fluid kinematic viscosity}
\defitem{${\bf{v}},{\bf{{v}}}^*$}\defterm{Dimensional and dimensionless fluid flow velocity vector}
\defitem{$p,{p}^*$}\defterm{Dimensional and dimensionless fluid flow pressure}
\defitem{${{{\bf{F}}_j}},{\bf{{f}}}_j^*$}\defterm{Dimensional and dimensionless fluid force on cylinder ${{C}_j}$}
\defitem{$h_j, \varphi_j$}\defterm{Magnitude and phase angle of ${\bf{{f}}}_j^*$}
\defitem{${\bf{U}},{\bf{{u}}}^*$}\defterm{Dimensional and dimensionless displacement vector of the cylinders}
\defitem{$U$}\defterm{Displacement module of the cylinders}
\defitem{${\rm{J}}_n,{\rm{Y}}_n$}\defterm{Bessel functions of the first and second kind}
\defitem{$\varepsilon$}\defterm{Dimensionless separation distance} 
\defitem{$KC$}\defterm{Keulegan-Carpenter number} 
\defitem{$Sk$}\defterm{Stokes number}
\defitem{$P/D$}\defterm{Pitch ratio}
\defitem{${\bf{e}}_x, {\bf{e}}_y$}\defterm{Cartesian basis vectors}
\defitem{$[M_j],[C_j]$}\defterm{Added mass and damping matrices}
\defitem{$m_{self}^{(j)}, c_{self}^{(j)}$}\defterm{Self-added mass and damping coefficients}
\defitem{$m_{cross}, c_{cross}$}\defterm{Cross-added mass and damping coefficient}
\defitem{$\widetilde{m}_{self}^{(j)}, \widetilde{c}_{self}^{(j)}$}\defterm{Asymptotic expansion of $m_{self}^{(j)}$ and $c_{self}^{(j)}$ as $Sk\to\infty$}
\defitem{$\widetilde{m}_{cross}, \widetilde{c}_{cross}$}\defterm{Asymptotic expansion of $m_{cross}$ and $c_{cross}$ as $Sk\to\infty$}
\defitem{$m_{self}^{(j)pot}, c_{self}^{(j)pot}$}\defterm{Inviscid limits of $m_{self}^{(j)}$ and $c_{self}^{(j)}$}
\defitem{$m_{cross}^{pot}, c_{cross}^{pot}$}\defterm{Inviscid limits of $m_{cross}$ and $c_{cross}$}
\defitem{$\iota$}\defterm{Relative deviation between theoretical/experimental and numerical predictions}
\end{deflist} 
\end{nomenclature}

\noindent These theoretical approaches have shown their efficiency in predicting the added-coefficients in some simple configurations, but extending to the industrial context remains delicate. As a consequence, engineers developed fast and robust numerical approaches, such as the immersed body method~\cite{gronski_2016}, the cut-cell method~\cite{cheny_2010} or the penalization method~\cite{Lagrange2020,mittal_2005,kadoch_2012, kolomenskiy_2011,schneider_2015,minguez_2008,nore_2018}. In the present work, we aim to analysis the efficiency of the Arbitrary Lagrangian-Eulerian method~\cite{donea2004arbitrary,fourestey2004second,koobus2000computation} implemented in the open-source code TrioCFD, developed by the CEA for the nuclear industry. TrioCFD is a C\texttt{++} object-oriented parallel software designed for calculations of unsteady laminar or turbulent fluid flows~\cite{Calvin2002}. The calculations are performed on structured (parallelepipeds) and non-structured (tetrahedrons) meshes of several millions of control volumes (hybrid finite volume element, see~\cite{Bieder2000a,Bieder2000b,Ducros2010}). The code structure is flexible, allowing the user to choose the discretization method, the convection and time schemes, as well as the turbulence model. The reader is referred to the TrioCFD webpage~\cite{TrioCFDwebpage} and~\cite{Angeli2015} for a detailed presentation of the code.\\

This article is organized as follows. Section~\ref{sec:2} presents the ALE method and the main equations solved by TrioCFD. In Section~\ref{sec:3}, we test the code, considering the vibration of two coaxial cylinders separated by a fluid layer. The added coefficients are extracted from the numerical predictions of the fluid forces, and the results are compared with those of the literature. In Section~\ref{sec:4}, we challenge the code, considering the vibration of a cylindrical tube located in the central position of a square tube array. Our numerical results for the fluid coefficients are compared with a phenomenological estimation used by engineers of the nuclear industry and with some experimental measurements that we have performed on a new experimental set-up built at CEA. Finally, Section~\ref{sec:5} summarizes our findings.

\section{Numerics}\label{sec:2}
The numerical simulations of the Navier-Stokes equations have been conducted on the 1.8.0 version of TrioCFD, a programmable CFD code based on the TRUST platform and a C++ language architecture, see~\cite{Angeli2015,Angeli2017}. This open source code allows the user to choose a wide range of options and parameters, among which the discretization schemes in space and time, the turbulence models (RANS or LES approaches) or the boundary conditions. Given its flexibility, the code is widely used in the nuclear industry for massive parallel and high performance calculations. 

In what follows, we briefly present the ALE method implemented in TrioCFD to solve fluid problems with moving boundaries.

\subsection{Principle of the ALE numerical method}
To determine the flow of a fluid, it is necessary to describe the kinematics of all its material particles throughout time. To do so, one can adopt either an Euler description of motion, in which a fluid particle is identified by its instantaneous position, or a Lagrange description of motion, in which a fluid particle is identified by its initial position. 
Both descriptions are totally equivalent, leading to different forms of the Navier-Stokes equations that can be discretized on a stationary mesh grid (Euler) or a mesh grid that follows the motion of the fluid particles (Lagrange). In both cases, the mesh grids do not account for the motion of the boundaries, which makes the numerical simulations of the related Navier-Stokes equations delicate. 

To overcome this problem, several approaches, such as the immersed boundary methods~\cite{puscas2015three, puscas2015time, puscas2015conservative}, or the ALE method~\cite{donea2004arbitrary, fourestey2004second, koobus2000computation} have been developed. 

In the ALE approach, a fluid particle is identified by its position relative to a frame moving with a nonuniform velocity ${\bf{v}}_{ALE}$. In this new frame of reference, the Navier-Stokes equations write
\begin{subequations}\label{eq:NS_ALE}
\begin{empheq}[left=\empheqlbrace]{align}
\nabla \cdot   \bf{v} &= 0, \\
\frac{\partial J {\bf{v}}}{\partial t}  &=  J \left(\nu\Delta {\bf{v}}   - \nabla \cdot  ( ({\bf{v}} - {\bf{v}}_{ALE}) \otimes {\bf{v}} )  - \frac{1}{\rho}\nabla p\right),
\end{empheq}
\end{subequations}
with $J$ the Jacobian of the transformation between the ALE and the Lagrange descriptions. The ALE method is actually a hybrid description between the Euler and the Lagrange descriptions, both of them corresponding to the particular cases ${\bf{v}}_{ALE}={\bf{0}}$ and ${\bf{v}}_{ALE}={\bf{v}}$, respectively. 

In the ALE framework, the choice of ${\bf{v}}_{ALE}$ is arbitrary as long as the deformation of the mesh grid remains under control. For moderate deformations, ${\bf{v}}_{ALE}$ is usually defined as the solution of an auxiliary Laplace problem, see~\cite{duarte2004arbitrary}:
\begin{subequations}\label{eq:mesh_motion}
\begin{empheq}[left=\empheqlbrace]{align}
\Delta {\bf{v}}_{ALE} &= {\bf{0}} \;\;\;\;\;\;\;\;\;\;\;\; \text{in the fluid domain,}\\
{\bf{v}}_{ALE} &= {\bf{v}}_{solid} \;\;\;\;\;\; \text{at a solid interface,}\\
{\bf{v}}_{ALE} &= {\bf{0}} \;\;\;\;\;\;\;\;\;\;\;\;\text{at a free surface,}
\end{empheq}
\end{subequations}
from which the kinematics of the mesh grid is updated, i.e. ${\bf{x}}^{new}={\bf{x}}^{old} + \Delta t{\bf{v}}_{ALE}$.

\subsection{Space and time discretizations}
\begin{figure}[ht!]
\begin{centering}
\begin{tabular}{ccc}
\begin{tikzpicture}[scale=0.6]

% place coordinates at the two initial vertices 
\coordinate  (A) at (2cm,6cm);
\coordinate  (B) at (3cm,0cm);
\coordinate  (C) at (8cm,2cm);
\coordinate  (xj) at (2.5cm,3cm);
\coordinate  (xi) at (5.5cm,1cm);
\coordinate  (xk) at (5cm,4cm);
\coordinate  (p) at (4.35cm,2.66cm);
\draw[thick] (A) -- (B) -- (C) --  (A);

\draw (xj) node[left] {$\mathbf{v}_h$};
\node at (xj) [rectangle,draw=black, ,fill=black] {};
\draw (5.5cm,0.9cm) node[below] {$\mathbf{v}_h$};
\node at (xi) [rectangle,draw=black, ,fill=black] {};
\draw (xk) node[right] {$\mathbf{v}_h$};
\node at (xk) [rectangle,draw=black, ,fill=black] {};

\draw [thick,dashed] (A) -- (xi);
\draw [thick,dashed] (B) -- (xk);
\draw [thick,dashed] (C) -- (xj);

\draw (4.35cm,2.55cm) node[left] {$p_h$};
\draw (p) node {$\bullet$};
\node at (A) [circle, ,draw=black] {};
\node at (B) [circle, ,draw=black] {};
\node at (C) [circle, ,draw=black] {};
\draw (A) node[left] {$p_h$};
\draw (B) node[left] {$p_h$};
\draw (C) node[right] {$p_h$};
\end{tikzpicture}
&  & 
\begin{tikzpicture}[scale=0.55]

\coordinate  (S1) at (3cm,0cm);
\coordinate  (S2) at (6cm,5cm);
\coordinate  (A) at (2cm,6cm);
\coordinate  (B) at (8cm,2cm);
\coordinate  (Cj) at (3.66cm,3.66cm);
\coordinate  (Ci) at (5.66cm,2.33cm);
\coordinate  (xj) at (2.5cm,3cm);
\coordinate  (xi) at (4.5cm,2.5cm);
\coordinate  (M) at (3.33cm,1.83cm);

\draw [thick] (S1) -- (S2) -- (A) --  (S1);
\draw[thick] (S1) -- (S2) -- (B) --  (S1);
\draw [thick,dashed] (S1) -- (Cj) -- (S2) --  (Ci) -- (S1);

%\draw  (3.53cm,2.64cm) node[right] {$\gamma_{ij}$};

\draw (Ci) node[below] {$G_i$};
\draw (Cj) node[above] {$G_j$};
\draw (Ci) node {$\bullet$};
\draw (Cj) node {$\bullet$};

\draw (xi) node[right] {$x_i$};
\draw (xj) node[left] {$x_j$};
\draw (xi) node {$\bullet$};
\draw (xj) node {$\bullet$};

\draw (S1) node[left] {$S_1$};
\draw (S2) node[right] {$S_2$};
\draw (S1) node {$\bullet$};
\draw (S2) node {$\bullet$};

\draw (6.6,2.2) node[right] {$K_i$};
\draw (2.1,5.3) node[right] {$K_j$};
\draw (3.6,3.1) node[right] {$w_i$};

\end{tikzpicture}\tabularnewline
(a) &  & (b)\tabularnewline
\end{tabular}
\par\end{centering}
\protect\caption{\label{fig:Freedom-degrees}(a) DoF of 2D element (black squares for velocity $\mathbf{v}_h$,  black dots and circles for pressure $p_h$). (b) Control volume $w_{i}$
between two triangles $K_{i}$ and $K_{j}$ of respective center $G_{i}$
and $G_{j}$.}
\end{figure}
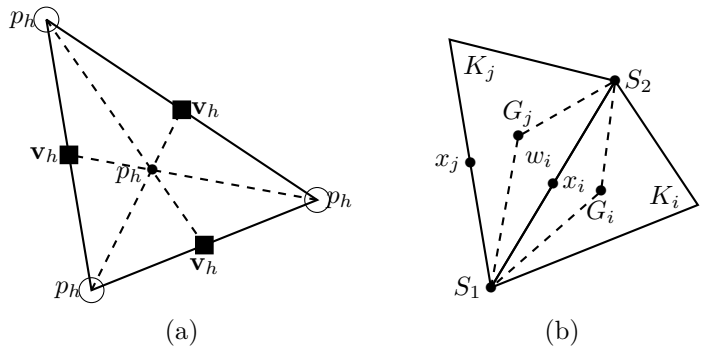

In case of unstructured meshes (triangles in 2D or tetrahedrons in 3D), TrioCFD uses Finite Volume-Elements (FVE) approach to solving discretized Navier-Stokes equations. These methods combine the Finite Elements (FE) method with the Finite Volumes (FV) method,  gathering the advantages of each approach for incompressible Navier-Stokes problems~\cite{Angeli2017}. 
The governing equations~\eqref{eq:NS_ALE} are solved through a staggered approaches: a primary grid for the discrete pressure $p_h$ and a face-based staggered dual grid for the discrete velocity $\b_v_h=(v_{x,h},v_{y,h})^{\operatorname{T}}$.
As shown in Fig.~\ref{fig:Freedom-degrees}, the velocity is evaluated on the centre of the 2D edges (or on the center of the 3D faces in 3D) while pressure has more degrees of freedom (DoF) and can be evaluated in the element gravity center and nodes.

As in Finite Volume approach, the local equations are integrated over the control volumes. The control volumes for the mass equation are the primal mesh cells whereas the dual mesh cells (denoted by $w$ hereafter) are the control volumes of momentum. The control volume $w$ associated to each face is obtained by joining the gravity centers of the two adjacent cells $G_i$ and $G_j$ with the vertices $S_1$ and $S_2$ of the shared face (see~\ref{fig:Freedom-degrees}(b)). The fluxes and the differential operators are computed by means of a FE formulation.

Hereafter, we recall the main ideas of the FVE method~\cite{Angeli2015,Fiorini2020} and for this purpose, in this section we only consider the spatial part of the systems~\eqref{eq:NS_ALE}. Also, in this paragraph, in order to simplify the presentation, we limit ourselves to the two-dimensional case. Let $\mathcal{T}_h$ be a triangulation of the domain and $K_j \in \mathcal{T}_h$ a triangle ($j=1,\ldots,N_T$). We denote with $\mathbf{x}_i$ the nodes, which are the middle points of the edges of the triangles and by $w_{i}$ ($i=1,\ldots,N_N$) their associated control volumes, see Fig.~\ref{fig:Freedom-degrees}(b). 
We introduce the following spaces:
\begin{subequations}
\begin{align}
Q_h &= \{q_h : \forall K_j \in \mathcal{T}_h,\ q_h|_{K_j} \in P_0(K_j)\},\\
W_h &= \{w_h \text{ continuous at }\mathbf{x}_i : \forall K_j \in \mathcal{T}_h,\ w_h|_{K_j} \in P_1(K_j)\},\\
\mathbf{W}_h &= \{\mathbf{w}_h=(w_{x,h}, w_{y,h})^{\operatorname{T}} : w_{x,h}, w_{y,h} \in W_h\} = W_h^2,
\end{align}
\end{subequations}
with $P_n$ a polynomial function of the $n$-$th$ order. 
The space $Q_h$ is spanned by the indicator functions of the triangles, $\chi_{K_j}$, and $W_h$ is spanned by $\phi_i(\mathbf{x})$, with $\phi_i \in W_h$ and $\phi_i(\mathbf{x}_j) = \delta_{ij}$.
We seek an approximate solution for the systems~\eqref{eq:NS_ALE}: $(\mathbf{v}_h, p_h) \in \mathbf{W}_h \times Q_h$.

In order to have a discrete formulation, we integrate the mass equation over the triangles $K_j$ and the momentum equation over the control volumes $\omega_i$ as:
\begin{subequations}\label{discr_integration}
\begin{empheq}[left=\empheqlbrace]{align}
-\int_{\partial \omega_i}(\nu\nabla \b_v_h - \dfrac{1}{\rho}p_h[I])\cdot\mathbf{n}d\sigma_{i} + \int_{\partial \omega_i}((\b_v_h -\b_v_{h, \text{ALE}}) \otimes \b_v_h)\cdot\mathbf{n}d\sigma_{i} &= {\bf{0}} \quad \forall i\in\left[1,\ldots,N_N\right], \\
\int_{\partial {K_j}} \b_v_h\cdot\mathbf{n}d\sigma_j &= 0 \quad \forall K_j\in \mathcal{T}_h,
\end{empheq}
\end{subequations}
with $d\sigma_{i}$ and $d\sigma_j$ a Lebesgue measure of $\partial\omega_i$ and $\partial K_j$, respectively, and $[I]$ the identity matrix. 

Then, one can expand $(\b_v_h,p_h)$ in the bases of the correspondent spaces as follows:
\begin{equation}
\begin{aligned}
\b_v_h (\mathbf{x}) &= \sum_{i=1}^{N_N} \b_v_h(\mathbf{x}_i)\phi_i(\mathbf{x}), \\
p_h (\mathbf{x}) &= \sum_{j=1}^{N_T} p_h(K_j) \chi_{K_j}.
\end{aligned}
\label{dev_discr_bases}
\end{equation}

Plugging (\ref{dev_discr_bases}) into (\ref{discr_integration}) for the linear terms and using the identity $(\mathbf{a}\otimes\mathbf{b})\cdot\mathbf{c} = \mathbf{a}(\mathbf{b}\cdot\mathbf{c})$ for the nonlinear terms, results in:
\begin{subequations}
\begin{empheq}[left=\empheqlbrace]{align}
\begin{split}
-\sum_{j=1}^{N_N}\int_{\partial \omega_i}\left(\nu\b_v_h(\mathbf{x}_j)\otimes\nabla\phi_j\right)\cdot\mathbf{n}d\sigma_i + \sum_{j=1}^{N_T} \dfrac{1}{\rho}p_h(K_j) \int_{\partial \omega_i}\chi_{K_j}\mathbf{n}d\sigma_i\;+\\
+ \int_{\partial \omega_i}\b_v_h(\b_v_h\cdot\mathbf{n})d\sigma_i
 - \int_{\partial \omega_i}\b_v_h(\b_v_{h, \text{ALE}}\cdot\mathbf{n})d\sigma_i  &={\bf{0}} \quad  \forall i\in\left[1,\ldots,N_N\right],\end{split}\\
\sum_{i=1}^{N_N} \b_v_h(\mathbf{x}_i) \cdot \int_{\partial {K_j}} \phi_i \mathbf{n}d\sigma_j  &=0  \quad \forall K_j\in \mathcal{T}_h.
\end{empheq}
\label{discr_integration_dev}
\end{subequations}
The unknown vectors are defined as follows:
\begin{equation}
\begin{aligned}
\V_h = \begin{bmatrix}
\mathcal{U}_h =[v_{x,h}(\mathbf{x}_i) ]_{i=1,\ldots,N_N} \\ \mathcal{V}_h = [v_{y,h}(\mathbf{x}_i)]_{i=1,\ldots,N_N}
\end{bmatrix},
\quad \mathbf{P}_h = \dfrac{1}{\rho}\left[p_h(K_j)\right]_{j=1,\ldots,N_T},
\end{aligned}
\end{equation}
and the following matrix elements:
\begin{equation*}
[\tilde{A}]_{ij} := -\int_{\partial\omega_i}\nu\nabla\phi_j\cdot\mathbf{n}d\sigma_i, \qquad
[D_\ell]_{ij} := \int_{\partial K_j}\phi_i n_\ell d\sigma_j, \qquad
[G_\ell]_{ij} := \int_{\partial\omega_i}\chi_{K_j} n_\ell d\sigma_i,
\end{equation*}
where the subscript $\ell$ indicates the components of the normal vector $\mathbf{n} = (n_x, n_y)$.

The convection term can be written as: 
\begin{equation}
\int_{\partial\omega_i} \b_v_h(\b_v_h\cdot\mathbf{n})d\sigma_i  = \int_{\partial\omega_i} \b_v_h\left(\displaystyle \sum_{j=1}^{N_N}\b_v_h(\mathbf{x}_j)\phi_j\cdot\mathbf{n}\right) d\sigma_i.
\label{eq:term_convectif}
\end{equation}
The $x$-projection of \eqref{eq:term_convectif} yields an equation for the first component $v_{x,h}$ of $\b_v_h$:
\begin{equation*}
\begin{split}
\int_{\partial\omega_i} v_{x,h}\left(\displaystyle \sum_{j=1}^{N_N}\b_v_h(\mathbf{x}_j)\phi_j\cdot\mathbf{n}\right)d\sigma_i = \sum_{j=1}^{N_N} v_{x,h}(\mathbf{x}_j) \int_{\partial\omega_i}v_{x,h}\phi_jn_x d\sigma_i + \sum_{j=1}^{N_N} v_{y,h}(\mathbf{x}_j) \int_{\partial\omega_i}v_{x,h}\phi_jn_y d\sigma_i.
\end{split}
\end{equation*}
We can define the following matrix elements:
\begin{equation*}
[L_x(\mathcal{U}_h)]_{ij} = \int_{\partial\omega_i}v_{x,h}\phi_jn_x d\sigma_i,\quad
[L_y(\mathcal{U}_h)]_{ij} = \int_{\partial\omega_i}v_{x,h}\phi_jn_y d\sigma_i.
\end{equation*}
The second component would give as a result $[L_x(\mathcal{V}_h)]_{ij}$ and $[L_y(\mathcal{V}_h)]_{ij}$. 
Finally, by introducing the following notation:
\begin{equation}
[A] = \left( \begin{matrix}
[\tilde{A}] & 0 \\ 0 & [\tilde{A}]
\end{matrix}\right),
\quad 
[D] = \left( \begin{matrix}
[D_x] & [D_y]
\end{matrix}\right),\quad 
[G] = \left( \begin{matrix}
[G_x] \\ [G_y]
\end{matrix}\right), \quad
[L(\V_h)] = \left(\begin{matrix}
[L_x(\mathcal{U}_h)] & [L_y(\mathcal{U}_h)] \\
[L_x(\mathcal{V}_h)] & [L_y(\mathcal{V}_h)] \\
\end{matrix}\right),
\end{equation}
the discrete system can be written in the compact form:
\begin{subequations}
\begin{empheq}[left=\empheqlbrace]{align}
 [A]\V_h - [L(\V_h)]\V_h +  [L(\V_h)]\V_{h,\text{ALE}} - [G] \mathbf{P}_h &= {\bf{0}},\\
[D] \V_h &= {\bf{0}}.
\end{empheq}
\end{subequations}

In the present work, the integral~\eqref{eq:term_convectif} is computed numerically by using MUSCL, i.e. a Monotone Upstream-Centred Scheme for Convective flows~\cite{van1979towards}, see~\ref{sec:muscl}.
Applying the numerical FVE discretization on Eq.~\eqref{eq:NS_ALE}, along with the Forward Euler scheme for the time discretization, results in the following discrete system:
\begin{subequations}\label{eq:DiscretizeNS}
\begin{empheq}[left=\empheqlbrace]{align}
[M] \frac{J^{n+1}\V_h^{n+1} - J^{n}\V_h^{n}}{\Delta t} & =J^{n+1} \left([A] \V_h^{n+1} - [L(\V_h^n)]\V_h^{n+1} +   [L(\V^n)]\V^{n+1}_{h, ALE} - [G] \mathbf{P}_h^{n+1}\right),\\
[D] \V_h^{n+1} &= {\bf{0}},
\end{empheq}
\end{subequations}
where $\Delta t$ is the time step (see~\ref{sec:D}), $[M]$ is the mass matrix operator, $[A]$ is the discrete diffusion operator, $[L(\V_h)]$ is the non-linear discrete convection operator, $[G]$ is the discrete gradient operator and $[D]$ is the discrete divergence operator, defined above.

In order to solve the velocity-pressure coupling, a multi-step (projection-correction) technique~\cite{chorin68, temam68} is employed, where an intermediate (predicted) velocity $\V_h^{*}$ is computed:
\begin{equation}
J^{n+1}\left( \frac{1}{\Delta t} [M] - [A] + [L(\V_h^n)]\right) \V_h^{*}  = \frac{1}{\Delta t} [M] J^{n}\V_h^n +  J^{n+1}[L(\V_h^n)]\V^{n}_{h, ALE} - J^{n+1} [G] \mathbf{P}_h^n,    
\label{eq:prediction}
\end{equation} 
and the mass conservation is then enforced by solving a Poisson equation for pressure:
\begin{equation}
[D] [M]^{-1} [G] \mathbf{P}_h^{'} =  \frac{1}{\Delta t} [D] \V^{*}_h.
\end{equation} 
Finally, the velocity is updated using the predicted velocity $\V_h^{*}$ and the pressure increment $\mathbf{P}_h^{'}$:
\begin{equation}
\V_h^{n+1} = \V_h^{*} - \Delta t [M]^{-1} [G] \mathbf{P}_h^{'},   \quad \mathbf{P}_h^{n+1} = \mathbf{P}_h^n +\mathbf{P}_h^{'}. 
\end{equation} 

In the present work, the linear systems~(\ref{eq:prediction}) is solved by the iterative solver GMRES from the PETSc library~\cite{saad1986gmres}.

%%%%%%%%%%%%%%%%%%%%%%%%%%%%%%%%%%%%%%%%%%%%%%%%%%%%%%%%%%%%%%%%%%%%%%%%%%%%%%%%%%%%%%%
%%%%%%%%%%%%%%%%%%%%%%%%%%%%%%%%%%%%%%%%%%%%%%%%%%%%%%%%%%%%%%%%%%%%%%%%%%%%%%%%%%%%%%%

\section{Study 1. Vibrations of two coaxial cylinders in a viscous fluid} \label{sec:3}
In this section, we test the capabilities of the code TrioCFD along with the ALE approach, considering the 2D case of two coaxial cylinders vibrating in a viscous fluid. The goal of this preliminary work is to introduce all the important concepts that will be used in the study of the vibrations of a tube in a square bundle, considered in \S\ref{sec:4}. 
In what follows, we introduce the problem, present the numerical setup and finally discuss our numerical results in comparison with the theoretical estimations of~\cite{Chen1976,Yeh1978}.  

\subsection{Presentation of the problem and governing equations}
\begin{figure}[H] 
	\centering
	\includegraphics[width=0.4\linewidth]{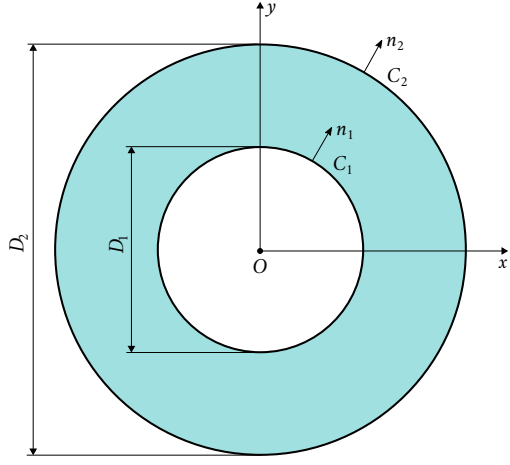}
	\caption{Configuration with two concentric cylinders. The fluid in the annulus region is homogeneous with mass density $\rho$ and kinematic viscosity $\nu$.}
	\label{fig:Initial_problem_1}
\end{figure}

Let $C_1$ and $C_2$ be two concentric cylinders with diameters $D_j$ and boundaries $\partial C_j$, $j=\{1,2\}$, see Fig.~\ref{fig:Initial_problem_1}. One cylinder, either $C_1$ or $C_2$, oscillates in the $\left(x,y\right)$ plane with a simple harmonic motion of angular frequency $\Omega$ and a displacement amplitude $U$. The fluid in the annulus region is Newtonian, homogeneous, with mass density $\rho$ and kinematic viscosity $\nu$.
The fluid flow generated by the oscillation of one cylinder is assumed as incompressible and two-dimensional. 

The Navier-Stokes equations and the boundary conditions for the fluid flow ($\mathbf{v},\;p$) are:
\begin{subequations}\label{eq:DimensionNS}
\begin{empheq}[left=\empheqlbrace]{align}
    \nabla \cdot \mathbf{v} &= 0, \label{eq:DimensionNS1}\\ 
	\dfrac{\partial \mathbf{v}}{\partial t} + (\mathbf{v} \cdot \nabla)\mathbf{v} + \dfrac{1}{\rho}\nabla p - \nu\Delta\mathbf{v} &= \mathbf{0}, \label{eq:eq:DimensionNS2}\\ 
	\mathbf{v} - \dfrac{d\mathbf{U}}{d t} &= \mathbf{0} \quad\text{on $\partial C_j$, $j=\{1,\;2\}$}. \label{eq:DimensionNS3}
\end{empheq}
\end{subequations}
The equation~\eqref{eq:DimensionNS3} ensures the continuity of velocities at the cylinder boundaries $\partial C_j$. The fluid force acting on $C_j$ is the sum of a pressure and a viscous term
\begin{equation}\label{eq:DimensionForce}
\mathbf{F}_j = - \int_{\partial C_{j}} p\mathbf{n}_j dL_j + \rho\nu \int_{\partial C_{j}} [\nabla \mathbf{v} + (\nabla \mathbf{v})^{\operatorname{T}}]\cdot\textbf{n}_j dL_j,
\end{equation}
where $\textbf{n}_j$ is the outward normal unit vector to $\partial C_j$, see Fig.~\ref{fig:Initial_problem_1}, and $dL_j$ is an infinitesimal line element of integration.

Picking $D_1/2$ and $\Omega^{-1}$ as a characteristic length and time, the dimensionless quantities $t^*$, $\textbf{u}^*$, $\textbf{v}^*$, $p^*$ and $\textbf{f}_j^*$ are defined as
\begin{equation}\label{eq:DimensionlessParameters1}
t =\Omega^{-1} t^*, \quad {\bf{U}} = U {\bf{u^*}}, \quad \textbf{v} = U\Omega\textbf{v}^*,\quad p = \rho U \dfrac{D_1}{2}\Omega^2p^*, \quad \textbf{F}_j = \rho U \bigg(\dfrac{D_1}{2}\Omega\bigg)^2\textbf{f}_j^*.
\end{equation}
To reduce the number of parameters of the problem, the following rescaled quantities are also defined
\begin{equation}\label{eq:DimensionlessParameters2}
\varepsilon = \dfrac{D_2}{D_1}, \quad KC = \dfrac{U}{D_1}, \quad Sk = \dfrac{D_1^2\left(\Omega /2\pi\right)}{\nu},
\end{equation}
as the dimensionless separation distance, the Keulegan–Carpenter number and the Stokes number, respectively.
Introducing~\eqref{eq:DimensionlessParameters1} in~\eqref{eq:DimensionNS} yields the dimensionless Navier–Stokes equations
\begin{subequations}\label{eq:DimensionlessNS}
   \begin{empheq}[left=\empheqlbrace]{align}
	\nabla \cdot \textbf{v}^* &= 0, \label{eq:DimensionlessNS1}\\ 
	\dfrac{\partial \textbf{v}^*}{\partial t^*} + 2 KC(\textbf{v}^* \cdot \nabla)\textbf{v}^* + \nabla p^* - \dfrac{2}{\pi Sk}\Delta\textbf{v}^* &= \textbf{0}, \label{eq:eq:DimensionlessNS2}\\
	\textbf{v}^* - \dfrac{d\textbf{u}^*}{d t^*} &= \textbf{0} \quad\text{on $\partial C_j$, $j=\{1,2\}$}. \label{eq:eq:DimensionlessNS3}
	\end{empheq}
\end{subequations}
The dimensionless fluid force acting on $C_j$ is
\begin{equation}\label{eq:DimensionlessForce}
\textbf{f}_j^* = - \int_{\partial C_{j}} p^*\textbf{n}_j dl_j + \dfrac{2}{\pi Sk}\int_{\partial C_{j}} [\nabla \textbf{v}^* + (\nabla \textbf{v}^*)^{\operatorname{T}}]\cdot\textbf{n}_j dl_j,
\end{equation}
with $dl_j = 2dL_j/D_1$.

For $KC<<1$, the dimensionless Navier-Stokes equations are linear. It follows that the fluid forces are linear combinations of the cylinder velocity $d{\bf{u}}^*/dt^*$ and acceleration $d^2{\bf{u}}^*/dt^{*2}$ 
\begin{equation}\label{eq:DimensionlessForces0}
\textbf{f}^*_j\\
= -\pi\left([M_{j}]
\frac{d^2{\bf{u}}^*}{dt^{*2}}
 + [C_{j}] \frac{d{\bf{u}}^*}{dt^{*}}
\right), 
\end{equation}
with
\begin{subequations}
\begin{align}
[M_{j}] &=
\begin{pmatrix}
m^{(j)}_{self}\\
m^{(j)}_{self}
\end{pmatrix}
\quad\;\;\;\;\text{and}\quad
 [C_{j}] = 
\begin{pmatrix}
c^{(j)}_{self}\\
c^{(j)}_{self}
\end{pmatrix}\quad\;\;\;\;\text{if $C_j$ is moving},\\
[M_{j}] &= 
\begin{pmatrix}
m_{cross}\\
-m_{cross}
\end{pmatrix}
\quad\text{and}\quad 
 [C_{j}] = 
\begin{pmatrix}
c_{cross}\\
-c_{cross}
\end{pmatrix}\quad\text{if $C_j$ is stationary}.
\end{align}
\end{subequations}
The self-added mass and damping coefficients $m_{self}^{(j)}$ and $c_{self}^{(j)}$ relate the fluid force on the moving cylinder to its own motion. The cross-added mass and damping coefficients $m_{cross}$ and $c_{cross}$ relate the fluid force on the stationary cylinder to the motion of the other cylinder.
The fluid added coefficients are functions of the dimensionless separation distance $\varepsilon$ and the Stokes number $Sk$.\\

For a harmonic displacement $\textbf{u}^*=\sin{(t^*)}\textbf{e}_x$, the dimensionless fluid forces~\eqref{eq:DimensionlessForces0} reduce to
\begin{subequations}\label{eq:DimensionlessForces}
\begin{align}
\mathbf{f}^*_j &= \pi \left(m_{self}^{(j)}\sin{(t^*)} - c_{self}^{(j)}\cos{(t^*)}\right)\mathbf{e}_x, \quad\; \text{if $C_j$ is moving},\\
\mathbf{f}^*_j &= \pi \left(m_{cross}\sin{(t^*)} - c_{cross}\cos{(t^*)}\right)\mathbf{e}_x, \quad \text{if $C_j$ is stationary},
\end{align}
\end{subequations}
or similarly $\mathbf{f}^*_j =h_j\sin\left(t^*+\varphi_j\right)\mathbf{e}_x$, with
\begin{subequations}\label{eq:DimensionlessMagnitudePhase}
\begin{align}
h_{j} &= \pi \sqrt{\left(m^{(j)}_{self}\right)^2 + \left(c^{(j)}_{self}\right)^2}\;\;\;\;\;\text{and} &\varphi_{j} &= -\arctan{\left(\frac{c^{(j)}_{self}}{m^{(j)}_{self}}\right)} \;\;\;\;\;\;\;\;\text{if $C_j$ is moving},\\ 
h_{j} &= \pi \sqrt{m_{cross}^2 + c_{cross}^2} \;\;\;\;\;\;\;\;\;\;\;\;\;\;\text{and}
&\varphi_{j} &= \pi-\arctan\left({\frac{c_{cross}}{m_{cross}}}\right) \;\;\;\text{if $C_j$ is stationary}. 
\end{align}
\end{subequations}
Yeh et al.~\cite{Yeh1978} derived some analytical expressions of the fluid added coefficients. Introducing $\Re$ and $\Im$ the real and imaginary part operators, these expressions write
\begin{subequations}\label{eq:exact_theory}
\begin{align}
m_{self}^{(1)} &= \Re(a_{11}), &m_{self}^{(2)} &= \varepsilon^2 \Re(a_{22}), &m_{cross} &= \varepsilon \Re(a_{12}),\\
c_{self}^{(1)} &= \Im(-a_{11}), &c_{self}^{(2)} &= \varepsilon^2 \Im(-a_{22}), &c_{cross} &= \varepsilon\Im(-a_{12}),
\end{align}
\end{subequations}
with $a_{11}=-(1+2a)$, $a_{22}=2a/\varepsilon$, $a_{12} = 1 - 2a/\varepsilon^2$ and 
\begin{equation}\label{eq:Determinant_Yin}
a = 
\begin{vmatrix}
1 & 1 & {\rm{J}}_1(\beta_1) & {\rm{Y}}_1(\beta_1) \\
0 & 1 & \varepsilon^{-1} {\rm{J}}_1(\beta_2) & \varepsilon^{-1} {\rm{Y}}_1(\beta_2) \\
2 & 2 & \beta_1 {\rm{J}}_0(\beta_1) & \beta_1 {\rm{Y}}_0(\beta_1)  \\
0 & 2 & \beta_1 {\rm{J}}_0(\beta_2) & \beta_1 {\rm{Y}}_0(\beta_2) 
\end{vmatrix}
\begin{vmatrix}
1 & 1 & {\rm{J}}_1(\beta_1) & {\rm{Y}}_1(\beta_1) \\
\varepsilon^{-2} & 1 & \varepsilon^{-1} {\rm{J}}_1(\beta_2) & \varepsilon^{-1} {\rm{Y}}_1(\beta_2) \\
0 & 2 & \beta_1 {\rm{J}}_0(\beta_1) & \beta_1 {\rm{Y}}_0(\beta_1)  \\
0 & 2 & \beta_1 {\rm{J}}_0(\beta_2) & \beta_1 {\rm{Y}}_0(\beta_2)
\end{vmatrix}^{-1}.
\end{equation}
In~\eqref{eq:Determinant_Yin}, ${\rm{J}}_n$ and ${\rm{Y}}_n$ are the Bessel functions of the first and second kind, with arguments $\beta_1=\left(1-{\rm{i}}\right)\sqrt{\pi}/2\sqrt{Sk}$ or $\beta_2=\varepsilon\beta_1$.

From an asymptotic expansion of these functions as $Sk \to \infty$, we show that $(m^{(j)}_{self},m_{cross},c^{(j)}_{self},c_{cross})$ are equivalent to $(\widetilde{m}_{self}^{(j)}, \widetilde{m}_{cross},\widetilde{c}_{self}^{(j)},\widetilde{c}_{cross})$ with
\begin{subequations}\label{eq:approximate_theory}
\begin{align}
\widetilde{m}_{self}^{(j)} &=  m_{self}^{(j)pot} + \dfrac{4}{\sqrt{\pi}}\dfrac{1}{\sqrt{Sk}}, &\widetilde{m}_{cross} &= m^{pot}_{cross} - \dfrac{4}{\sqrt{\pi}}\dfrac{1}{\sqrt{Sk}},\\
\widetilde{c}_{self}^{(j)} &= \dfrac{4}{\sqrt{\pi}}\dfrac{1}{\sqrt{Sk}}\dfrac{\varepsilon^4 + \varepsilon}{(\varepsilon^2 - 1)^2}, &\widetilde{c}_{cross} &= -\dfrac{4}{\sqrt{\pi}}\dfrac{1}{\sqrt{Sk}}\dfrac{\varepsilon^4 + \varepsilon}{(\varepsilon^2 - 1)^2},
\end{align}
\end{subequations}
and
\begin{equation}\label{eq:PC}
m_{self}^{(1)pot} = \dfrac{\varepsilon^2 + 1}{\varepsilon^2 - 1}, \quad m_{self}^{(2)pot} = \dfrac{\varepsilon^2(\varepsilon^2 + 1)}{\varepsilon^2 - 1}, \quad m_{cross}^{pot} = -\dfrac{2\varepsilon^2}{\varepsilon^2 - 1}.
\end{equation}
The terms $m_{self}^{(j)pot}$ and $m^{pot}_{cross}$ are the inviscid limits of $m_{self}^{(j)}$ and $m_{cross}$ as $Sk \to \infty$. These terms can also be obtained from a potential theory in which the fluid forces are only due to the pressure field. Note that $c^{(j)pot}_{self} = c^{pot}_{cross} = 0$ as there are not damping effects in the inviscid framework.\\

In \S\ref{Study1_results} the theoretical estimations~\eqref{eq:exact_theory} and~\eqref{eq:approximate_theory} are compared with the numerical predictions performed with TrioCFD. 
Numerically, the fluid forces acting on $\partial C_j$ are computed by the sum of the pressure and viscous terms in Eq.~\eqref{eq:DimensionForce} given by TrioCFD. To extract the added coefficients from the numerical simulations of the dimensionless fluid forces, we introduce the Fourier inner product over five periods
\begin{equation}\label{eq:Fourier}
    \langle f(t^*),g(t^*) \rangle = \dfrac{1}{5\pi}\int_{0}^{10\pi}f(t^*)g(t^*)dt^*.
\end{equation}
From~\eqref{eq:DimensionlessForces}, it follows that the added-coefficients are
\begin{subequations}\label{eq:numerical_coefficient}
\begin{align}
m^{(j)}_{self} &= \dfrac{\langle\sin(t^*), \textbf{f}^*_{j}(t^*) \cdot \textbf{e}_x\rangle}{\pi}\;\;\text{and}\;\;c^{(j)}_{self} = -\dfrac{\langle\cos(t^*),\textbf{f}^*_{j}(t^*) \cdot \textbf{e}_x\rangle}{\pi}\;\;\;\text{if $C_j$ is moving},\\
m_{cross} &= \dfrac{\langle\sin(t^*), \textbf{f}^*_{j}(t^*) \cdot \textbf{e}_x\rangle}{\pi}\;\;\text{and}\;\;
c_{cross} = -\dfrac{\langle\cos(t^*), \textbf{f}^*_{j}(t^*) \cdot \textbf{e}_x\rangle}{\pi}\;\;\text{if $C_j$ is stationary}.
\end{align} 
\end{subequations}

\subsection{Numerical setup}
To discretize the fluid domain, we use an unstructured grid of triangles generated by the Gmsh platform, see~\cite{Geuzaine2009}. Due to its adaptive and automatic algorithm, Gmsh makes it possible to choose between two different local sizes for the 2D mesh: a small local size, $lc_{fine}$, for elements close to the moving cylinder and a large local size, $lc$, for elements close to the stationary cylinder. By this way, a refined mesh is used in the regions with large gradient fields whereas a loose mesh is used in the areas with low gradient fields. In~\ref{sec:effect_mesh_size}, a mesh sensitivity analysis clearly shows a convergence of the mass coefficients as $l_c$ and $lc_{fine}$ are changed. The convergence of the damping coefficients is less obvious, especially for high values of $Sk$. Physically, this is related to the thickness of the boundary layer, which tends to zero as $Sk$ increases. It follows that a finer mesh is required close to a cylinder boundary to account for the thickness of the boundary layer and obtain an accurate estimation of the damping terms. In this work, a compromise between the time of calculation and the precision (throughout of this work, we indicate the relative deviations of our numerical results compared to some theoretical or experimental references) is made to choose adequate values for $lc$ and $lc_{fine}$.    

\subsection{Results and discussion}\label{Study1_results}
%\subsubsection{Study 1. Case 1. $C_1$ is moving, $C_2$ is stationary}
The inner cylinder is imposed a sinusoidal displacement in the $x$-direction for five periods of time, i.e. $\textbf{u}^* = \sin(t^*)\textbf{e}_x$ with $t^* \in \{0,10\pi\}$. The numerical simulations are performed for $\varepsilon \in \{1.25,1.5,2\}$, $Sk \in \{10^1,10^2,10^3,10^4\}$ and $KC = 10^{-2}$.\\

\begin{figure}[!ht]
	\centering
	\includegraphics[width = 0.9\linewidth]{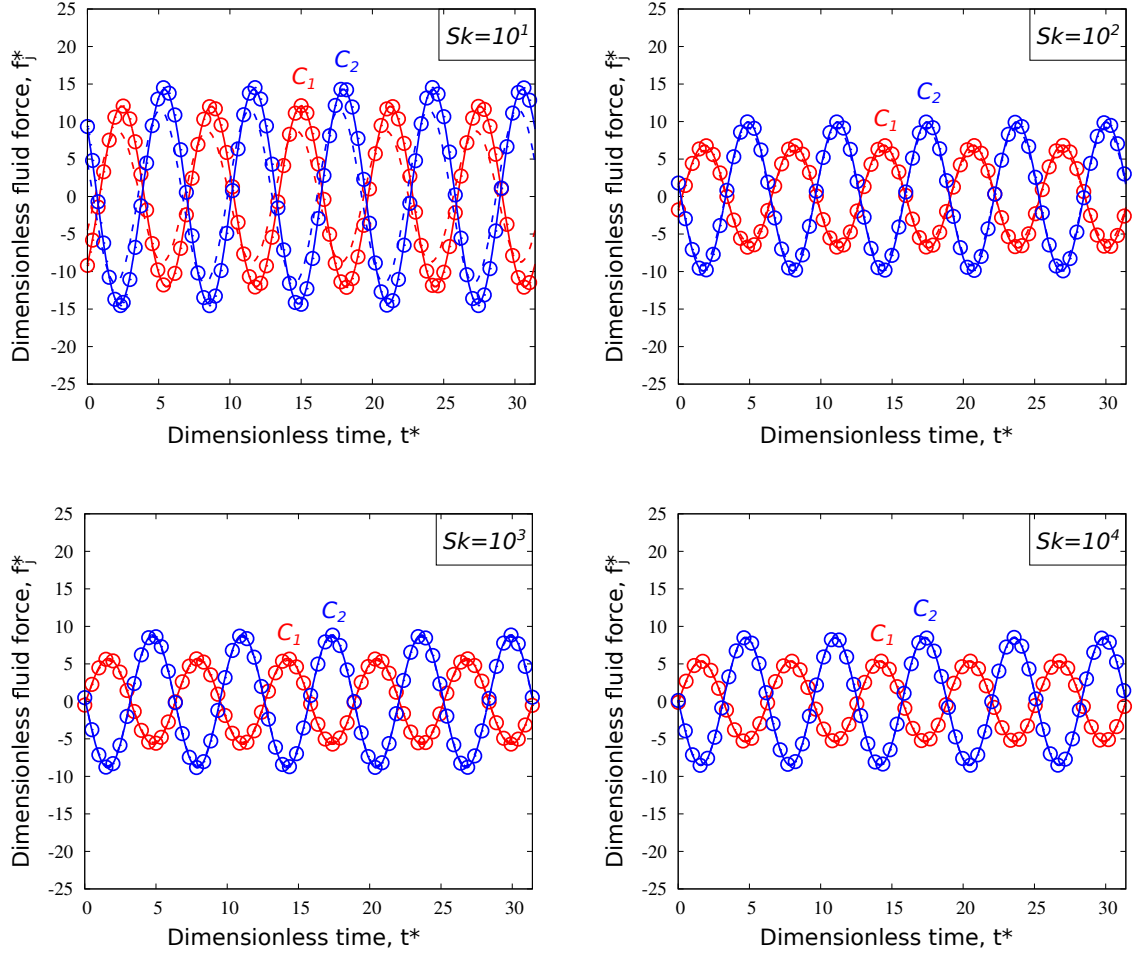}
	\caption{Study 1. Time evolution of the dimensionless fluid forces, for $Sk=\{10^1,10^2,10^3,10^4\}$, $\varepsilon = 2$ and $KC = 10^{-2}$. The red and blue solid lines correspond to the theoretical estimations of~\cite{Yeh1978}, see Eq.~\eqref{eq:exact_theory}. The dashed lines (indistinguishable from the solid lines for $Sk>10^2$) correspond to the theoretical asymptotic expansion $Sk\to\infty$, see  Eq.~\eqref{eq:approximate_theory}. The symbols correspond to the numerical predictions.}
	\label{fig:dimensionless_force_Sk}
\end{figure}

\begin{figure}[!ht]
	\centering
	\includegraphics[width = 0.9\linewidth]{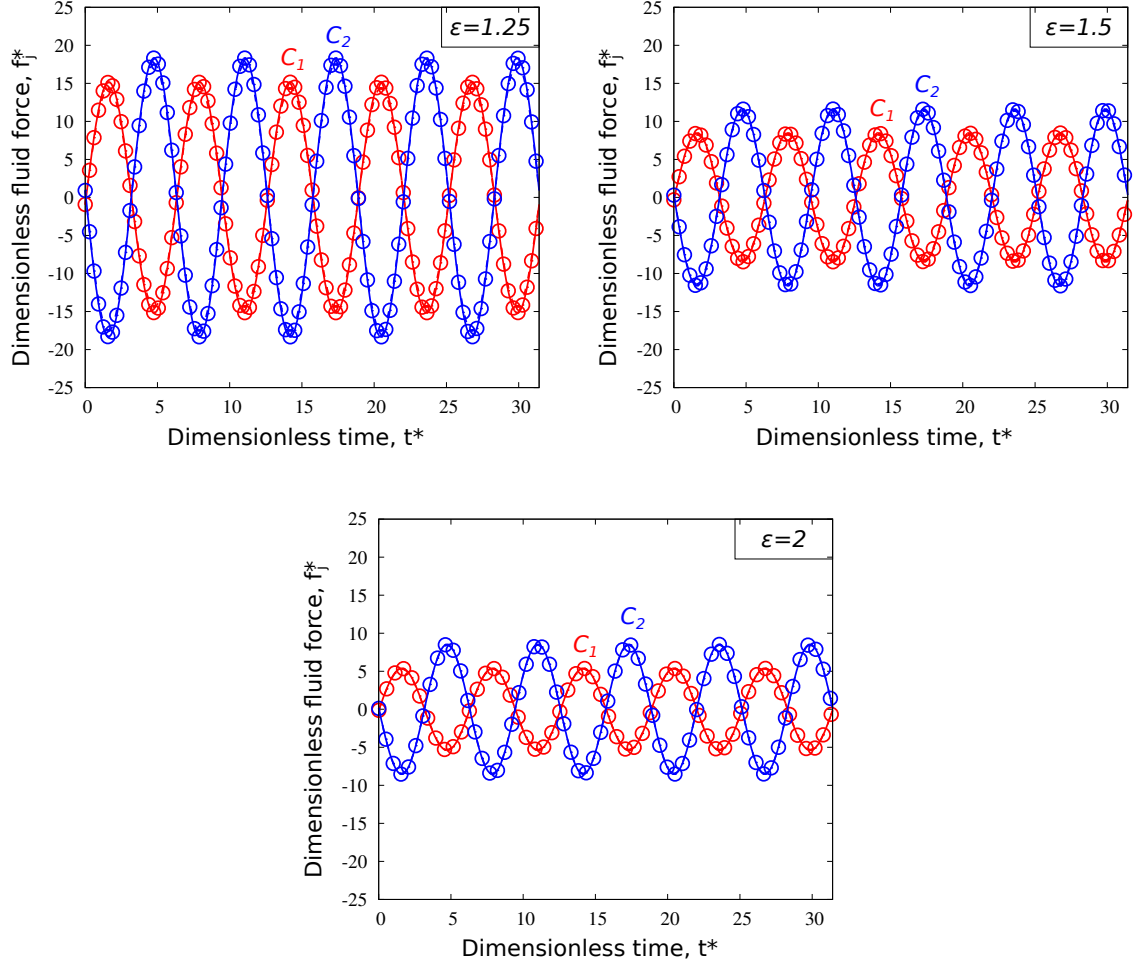}
	\caption{Study 1. Time evolution of the dimensionless fluid forces, for $\varepsilon=\{1.25,1.5,2\}$, $Sk = 10^{4}$ and $KC = 10^{-2}$. The red and blue solid lines correspond to the theoretical estimations of~\cite{Yeh1978}, see Eq.~\eqref{eq:exact_theory}. The symbols correspond to the numerical predictions.}
	\label{fig:dimensionless_force_eps}
\end{figure}

\begin{figure}[!ht]
	\centering
	\includegraphics[width = 0.9\linewidth]{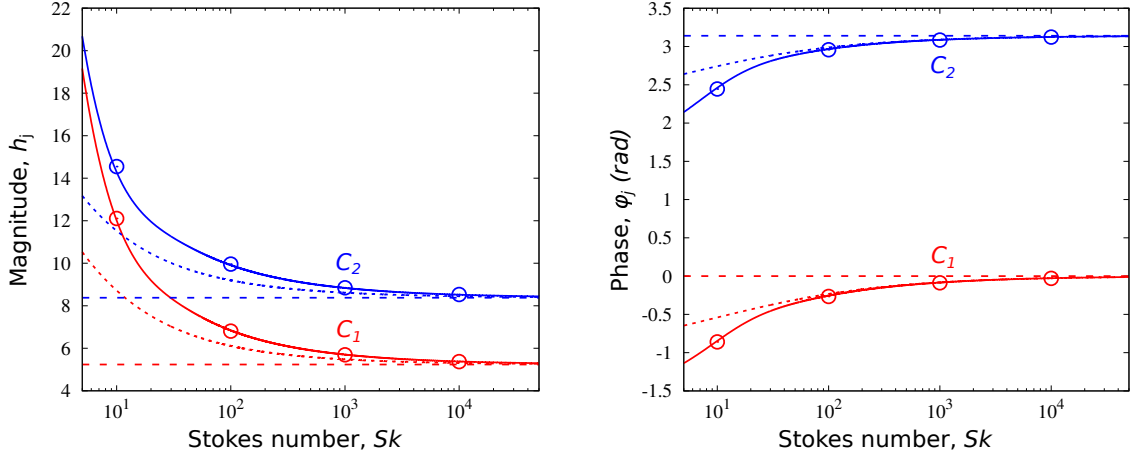}
	\caption{Study 1. Evolution of the magnitudes $h_j$ and phases $\varphi_j$ with the Stokes number $Sk$, for $\varepsilon = 2$ and $KC = 10^{-2}$. The red and blue solid lines correspond to the theoretical estimations of~\cite{Yeh1978}, see Eq.~\eqref{eq:exact_theory}. The dashed lines correspond to the theoretical asymptotic expansion $Sk\to\infty$, see  Eq.~\eqref{eq:approximate_theory}. The horizontal dashed lines correspond to the theoretical inviscid limits as $Sk \to \infty$. The symbols correspond to the numerical predictions.}
	\label{fig:amplitude_phase_Sk}
\end{figure}

\begin{figure}[!ht]
	\centering
	\includegraphics[width = 0.9\linewidth]{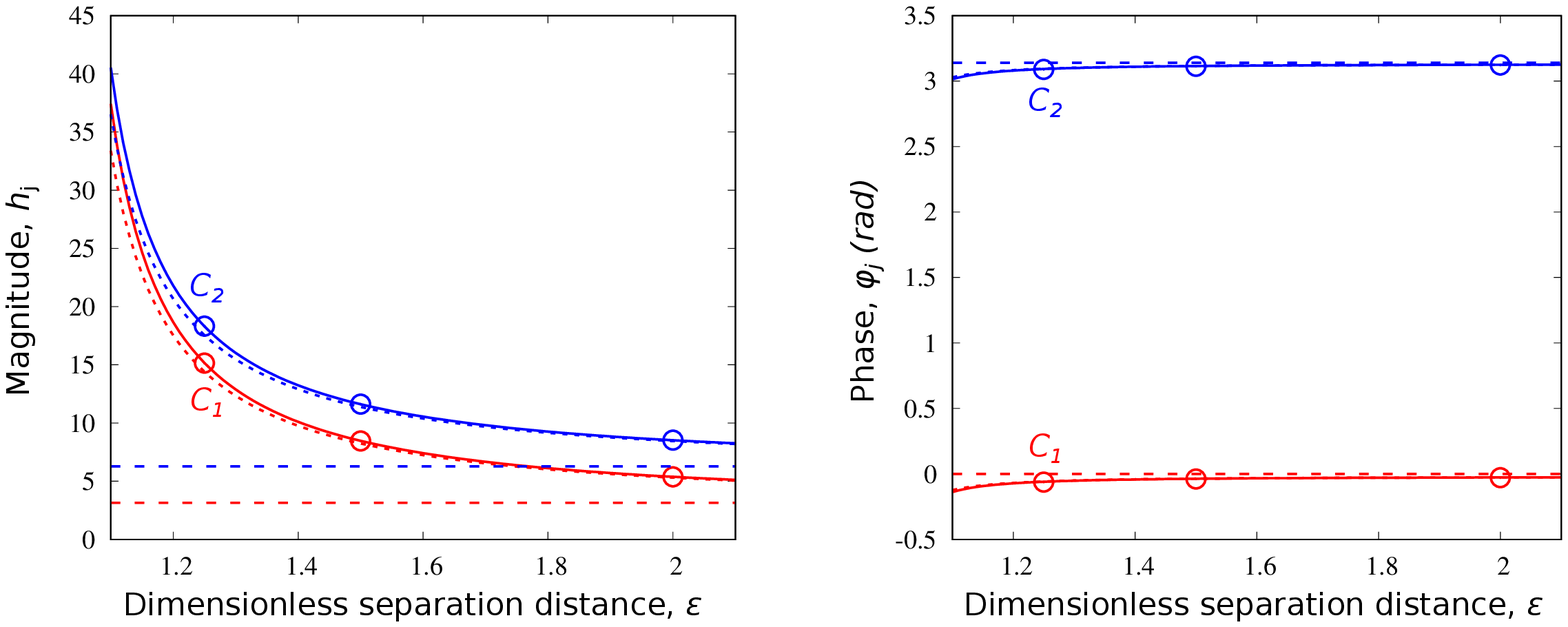}
	\caption{Study 1. Evolution of the magnitudes $h_j$ and phases $\varphi_j$ with the dimensionless separation distance $\varepsilon$, for $Sk = 10^{4}$ and $KC = 10^{-2}$. The red and blue solid lines correspond to the theoretical estimations of~\cite{Yeh1978}, see Eq.~\eqref{eq:exact_theory}. The dashed lines correspond to the theoretical asymptotic expansion $Sk\to\infty$, see  Eq.~\eqref{eq:approximate_theory}. The horizontal dashed lines correspond to the theoretical isolated cylinder limits as $\varepsilon \to \infty$. The symbols correspond to the numerical predictions.}
	\label{fig:amplitude_phase_eps}
\end{figure}

The time evolution of the dimensionless fluid forces is represented in Figs.~\ref{fig:dimensionless_force_Sk} and~\ref{fig:dimensionless_force_eps}. A perfect agreement is obtained between theoretical estimations and numerical predictions. The fluid forces are sinusoidal functions in phase opposition. In Figs.~\ref{fig:amplitude_phase_Sk} and~\ref{fig:amplitude_phase_eps}, we show that the amplitudes $h_j$ (resp. phases $\varphi_j$) decrease (resp. increase) with both $Sk$ and $\varepsilon$, recovering the asymptotic limits as $Sk \to \infty$ (inviscid limit) and as $\varepsilon \to \infty$ (isolated cylinder limit).\\

%%%%%%%%%%%%%%%%%%%%%%%%%%%%%%%%%%%%%%%%%%%%%%%%%%%%%%%%%%
% Case 1. Evolution of forces with Sk
%%%%%%%%%%%%%%%%%%%%%%%%%%%%%%%%%%%%%%%%%%%%%%%%%%%%%%%%%%
%%%%%%%%%%%%%%%%%%%%%%%%%%%%%%%%%%%%%%%%%%%%%%%%%%%%%%%
% Case 1. Evolution of forces with Eps
%%%%%%%%%%%%%%%%%%%%%%%%%%%%%%%%%%%%%%%%%%%%%%%%%%%%%%%%%%
%%%%%%%%%%%%%%%%%%%%%%%%%%%%%%%%%%%%%%%%%%%%%%%%%%%%%%%%%%
% Case 1. Evolution of magnitude and phase with Sk
%%%%%%%%%%%%%%%%%%%%%%%%%%%%%%%%%%%%%%%%%%%%%%%%%%%%%%%%%%
%%%%%%%%%%%%%%%%%%%%%%%%%%%%%%%%%%%%%%%%%%%%%%%%%%%%%%%%%%
% Case 1. Evolution of magnitude and phase with Eps
%%%%%%%%%%%%%%%%%%%%%%%%%%%%%%%%%%%%%%%%%%%%%%%%%%%%%%%%%%

\begin{figure}[!ht]
	\centering
	\includegraphics[width = 0.9\linewidth]{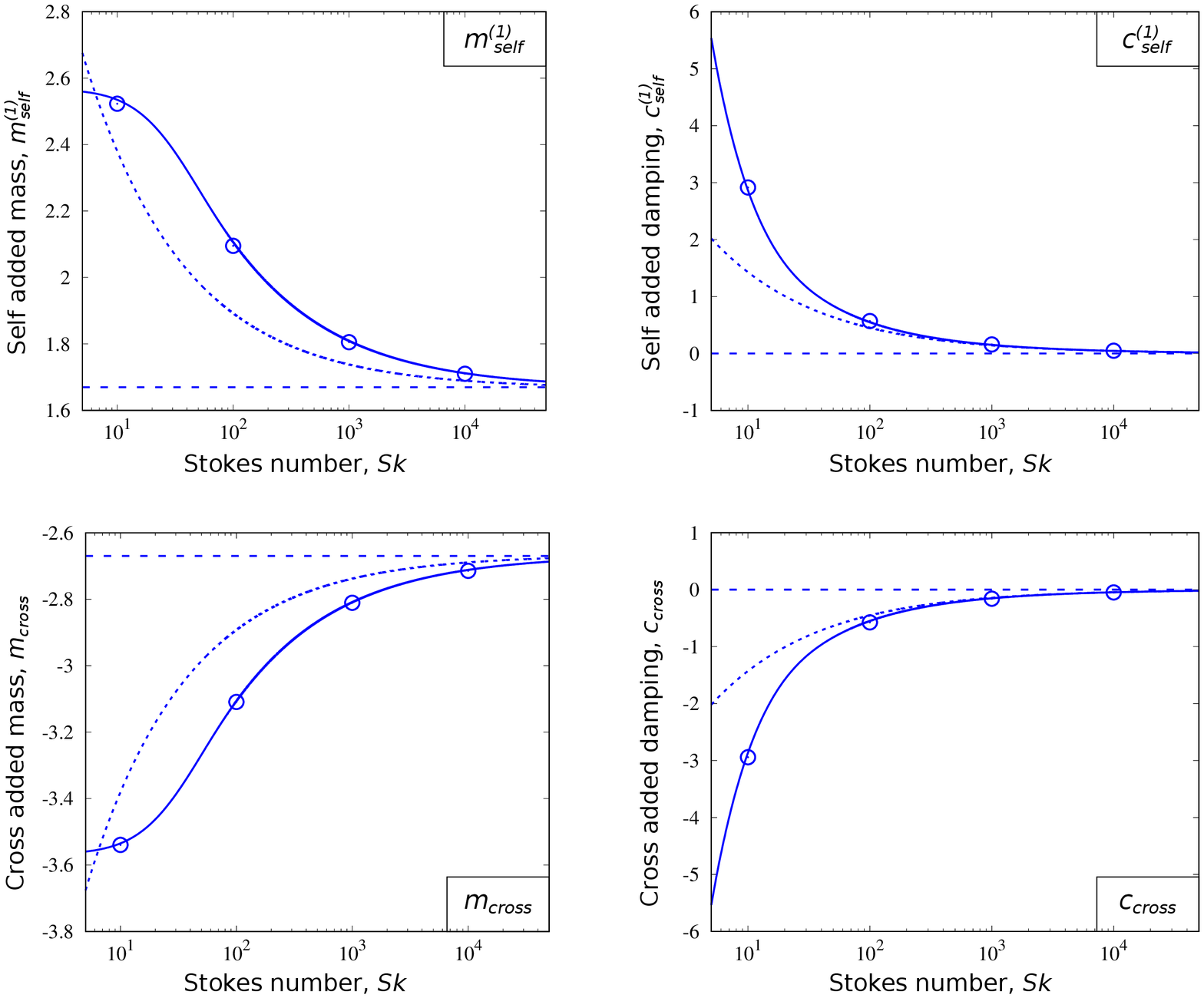}
	\caption{Study 1. Evolution of the added coefficients with the Stokes numbers $Sk$, for $\varepsilon = 2$ and $KC = 10^{-2}$. The solid lines refer to the theoretical estimations of~\cite{Yeh1978}, see Eq.~\eqref{eq:exact_theory}. The dashed lines correspond to the theoretical asymptotic expansion $Sk\to\infty$, see  Eq.~\eqref{eq:approximate_theory}. The horizontal dashed lines correspond to the theoretical inviscid limits as $Sk \to \infty$. The symbols correspond to the numerical predictions.}
	\label{fig:coefficients_Sk}
\end{figure}

\begin{figure}[!ht]
	\centering
	\includegraphics[width = 0.9\linewidth]{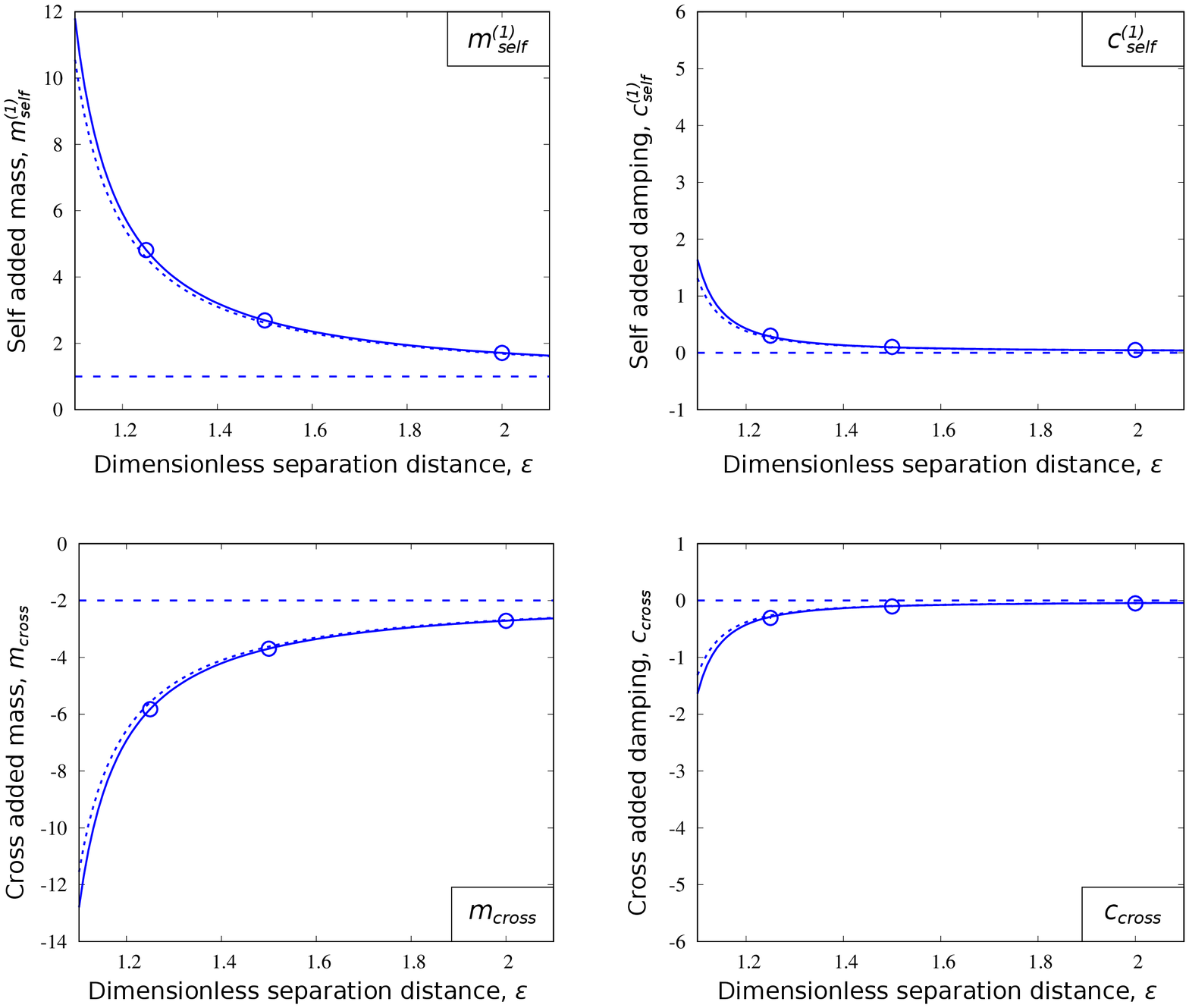}
	\caption{Study 1. Evolution of the added coefficients with the dimensionless separation distances $\varepsilon$, for $Sk = 10^{4}$ and $KC = 10^{-2}$. The solid lines refer to the theoretical estimations of~\cite{Yeh1978}, see Eq.~\eqref{eq:exact_theory}. The dashed lines correspond to the theoretical asymptotic expansion $Sk\to\infty$, see  Eq.~\eqref{eq:approximate_theory}. The horizontal dashed lines correspond to the theoretical isolated cylinder limits as $\varepsilon \to \infty$. The symbols correspond to the numerical predictions.}
	\label{fig:coefficients_eps}
\end{figure}

The evolution of the added coefficients is depicted in Figs.~\ref{fig:coefficients_Sk} and~\ref{fig:coefficients_eps}. The numerical predictions are in very good agreement with the theoretical estimations, even if a tiny difference is observed in the range of low Stokes numbers in which the theoretical estimations reach their limit of validity. Still, we confirm that the self (resp. cross) added coefficients decrease (resp. increase) with both $Sk$ and $\varepsilon$, recovering the asymptotic limits as $Sk \to \infty$ (inviscid limit) and as $\varepsilon \to \infty$ (isolated cylinder limit).\\

%%%%%%%%%%%%%%%%%%%%%%%%%%%%%%%%%%%%%%%%%%%%%%%%%%%%%%%%%%
% Case 1. Evolution of added coefficients with Sk
%%%%%%%%%%%%%%%%%%%%%%%%%%%%%%%%%%%%%%%%%%%%%%%%%%%%%%%%%%
%%%%%%%%%%%%%%%%%%%%%%%%%%%%%%%%%%%%%%%%%%%%%%%%%%%%%%%%%%
% Case 1. Evolution of added coefficients with Eps
%%%%%%%%%%%%%%%%%%%%%%%%%%%%%%%%%%%%%%%%%%%%%%%%%%%%%%%%%%

\begin{figure}[!ht]
	\centering
	\includegraphics[width = 0.9\linewidth]{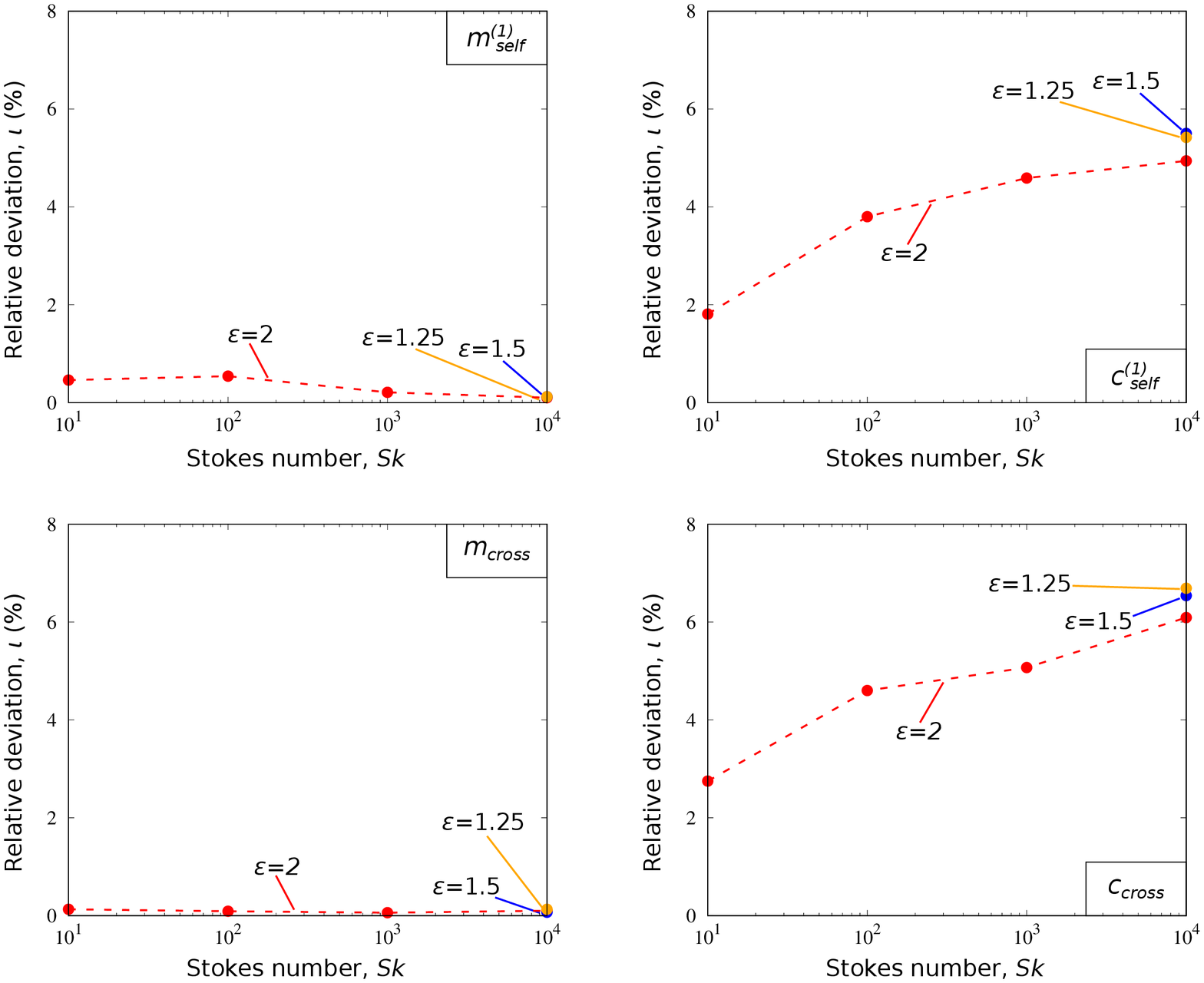}
	\caption{Study 1. Evolution of the relative deviation, $\iota$ with the Stokes number, $Sk$ for different value of $\varepsilon$ and $KC=10^{-2}$. The red dotted lines with points refer to $\varepsilon=2$, the blue points refer to $\varepsilon=1.5$ while the orange points refer to $\varepsilon=1.25$.}
	\label{fig:iota_Sk_eps}
\end{figure}

We introduce the quantity $\iota$, defined as the relative deviation between the numerical and the exact theoretical predictions~\cite{Yeh1978} of some quantity $Q$ : $\iota=|Q_{\rm{th.}}-Q_{\rm{num.}}|/|Q_{\rm{th.}}|$. The Fig.~\ref{fig:iota_Sk_eps} and the Tabs.~\ref{tab:Comparison_theory_numerics_Sk},~\ref{tab:Comparison_theory_numerics_eps} in~\ref{sec:AppendixA_numerics_vs_theory} show that $\iota$ is very low for the added mass coefficients and it becomes greater for the added damping coefficients as $Sk$ increases but seems to be weakly influenced by $\varepsilon$. In fact, for large values of $Sk$, the damping coefficients become very small and even small differences from the theoretical value produce significant deviations. Physically, this is related to the thickness of the boundary layer, which tends to zero: to reduce $\iota$ for the damping coefficients a finer mesh at the cylinder boundaries would be required. We shall note that the theoretical approach~\cite{Chen1976,Yeh1978} is fully linear since the convective term $2KC{{\left({\bf{{v}}^*} \cdot \nabla\right) {\bf{{v}}^*} }}$ of the Navier-Stokes equation~\eqref{eq:eq:DimensionlessNS2} is neglected. In the numerical simulations, the nonlinear convective term is retained through a small but nonzero Keulegan-Carpenter number $KC=10^{-2}$. However, this difference might slightly affect the deviation between the theoretical and numerical results. In any case, the relative deviation for $m^{(1)}_{self}$ (resp. $m_{cross}$) is smaller than $\iota \le 0.6\%$ (resp. $\iota\le0.2\%$) while the relative deviation for $c^{(1)}_{self}$ (resp. $c_{cross}$) is more pronounced, with $\iota \le 5.5\%$ (resp. $\iota \le 6.7\%$).\\

%%%%%%%%%%%%%%%%%%%%%%%%%%%%%%%%%%%%%%%%%%%%%%%%%%%%%%%%%%
% Case 1. Evolution of iota with Sk and Eps
%%%%%%%%%%%%%%%%%%%%%%%%%%%%%%%%%%%%%%%%%%%%%%%%%%%%%%%%%%

A further analysis is carried out to check the scale invariance of the numerical results obtained for $\varepsilon = 2$, $Sk = 10^4$ and $KC = 10^{-2}$. In the numerical simulation, the dimensional quantities were : $D_1 = 1$, $D_2 = 2$, $U=10^{-2}$ unit of length and $\Omega=0.0633$ unit of frequency. In this new case study,  we set $D_1 = 0.0316$, $D_2 = 0.0632$, $U=0.0316 \times 10^{-2}$ unit of length and $\Omega=63.3$ unit of frequency, such that they are self similar configurations in terms of dimensionless numbers: $\varepsilon = 2$, $Sk = 10^4$ and $KC = 10^{-2}$. The time evolution of the dimensionless fluid forces is represented in Fig.~\ref{fig:dimensionless_force_inv}, for both cases. As expected, the new numerical predictions $\left(m_{self}^{(1)},~m_{cross},~c_{self}^{(1)},~c_{cross}\right) = \left(1.71,~-2.71,~0.0496,~-0.0510\right)$ are strictly identical to those obtained for the first case study $\left(m_{self}^{(1)},~m_{cross},~c_{self}^{(1)},~c_{cross}\right) = \left(1.71,~-2.71,~0.0483,~-0.0488\right)$, confirming that the dimensionless added coefficients are functions of $Sk$, $\varepsilon$ and $KC$.\\

Finally, we check the symmetry of the fluid added mass and damping matrices for $\varepsilon = 2$, $Sk = 10^4$ and $KC = 10^{-2}$. Therefore, the outer cylinder is imposed a sinusoidal displacement in the $x$-direction for five periods of time, i.e. $\textbf{u}^* = \sin(t^*)\textbf{e}_x$ with $t^* \in \{0, 10\pi\}$. The time evolution of the dimensionless fluid forces is represented in Fig.~\ref{fig:dimensionless_force_outer}(a). The numerical predictions for the added coefficients are $\left(m_{self}^{(2)},~m_{cross},~c_{self}^{(2)},~c_{cross}\right) = \left(6.71,~-2.70,~0.0519,~-0.0488\right)$ whereas the theoretical estimations are $\left(m_{self}^{(2)},~m_{cross},~c_{self}^{(2)},~c_{cross}\right) = \left(6.71,~-2.71,~0.0460,~-0.0460\right)$. It follows that the relative deviation between the two approaches is $\iota\leq 12.8\%$. We attribute this deviation to the mesh refinement and the larger extension of the surface on which the ALE module acts, compared to the case in which the inner cylinder moves. In fact, in the ALE approach we need to solve the auxiliary Laplace problem~\eqref{eq:mesh_motion} to compute the mesh velocity, with its own approximations and therefore with an additional numerical error. We shall note that a finer mesh would have reduced this error and normally we would not have observed such differences. 

As expected, the numerical simulations confirm that the cross added coefficients do not depend on which cylinder is moving. Indeed, for $\varepsilon = 2$, $Sk = 10^4$ and $KC = 10^{-2}$, when $C_1$ moves we obtained $\left(m_{cross},~c_{cross}\right)=\left(-2.71,~-0.0488\right)$ whereas when $C_2$ moves we have $\left(m_{cross},~c_{cross}\right)=\left(-2.70,~-0.0488\right)$. In other words, the dimensionless fluid force acting on $C_1$ as $C_2$ moves is the same as the dimensionless fluid force acting on $C_2$ as $C_1$ moves, as shown in Fig.~\ref{fig:dimensionless_force_outer}(b).\\

\begin{figure}[!ht]
\centering
	\includegraphics[scale=0.22]{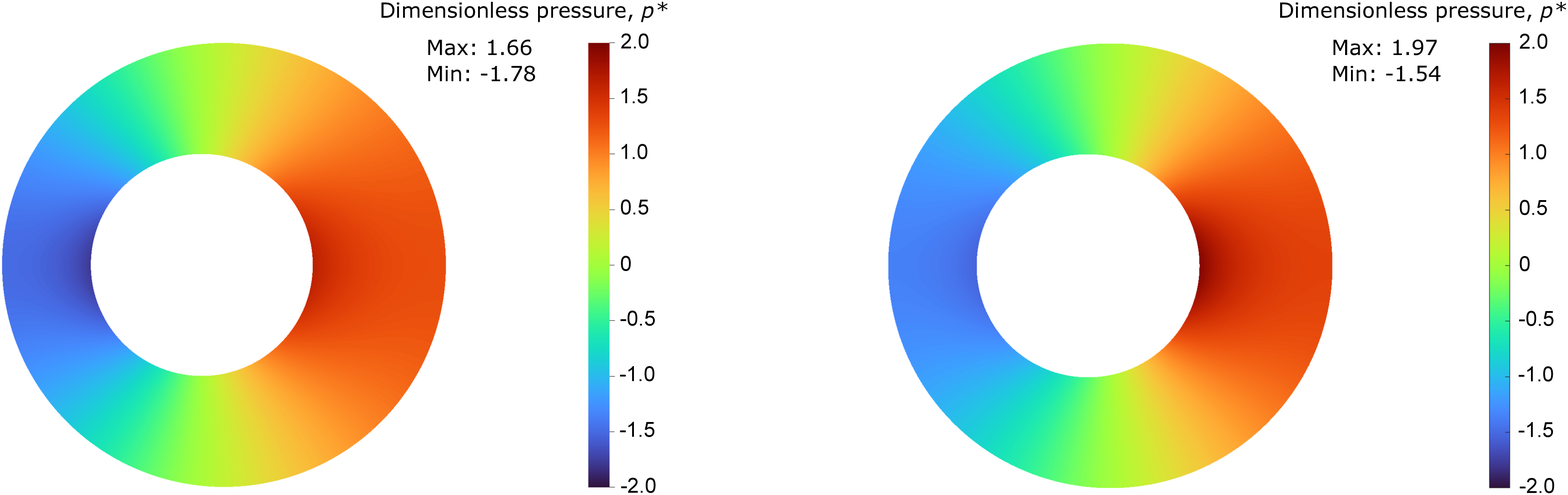}
	\caption{Study 1. Dimensionless pressure field, for $Sk = 10^{4}$, $\varepsilon = 2$ and $KC = 10^{-1}$. The right side figure corresponds to the theoretical estimation of \cite{Chen1976}. The left side figure corresponds to the TrioCFD prediction (the time evolution of the pressure field can be observed in the corresponding movie \cite{Pressure}).}
	\label{fig:pressure_field_TwoCylinders}
\end{figure}

As a conclusion to study 1, we show in Fig.~\ref{fig:pressure_field_TwoCylinders} the pressure distribution in the fluid layer when $C_1$ is imposed a sinusoidal displacement in the $x$-direction and $C_2$ is stationary, for $Sk = 10^{4}$, $\varepsilon = 2$ and $KC = 10^{-1}$. The numerical and theoretical predictions are in very good agreement as the two snapshots look very similar. Note that these snapshots are taken at a time when $C_1$ has a positive acceleration. As expected from the inertia effect of the added mass, the fluid pushes $C_1$ in a direction opposite to its acceleration vector, by creating a negative (resp. positive) pressure on the left (resp. right) side of $C_1$.\\
%%%%%%%%%%%%%%%%%%%%%%%%%%%%%%%%%%%%%%%%%%%%%%%%%%%%%%%%%%
%Study 1. Case 1. Dimensionless pressure field
%%%%%%%%%%%%%%%%%%%%%%%%%%%%%%%%%%%%%%%%%%%%%%%%%%%%%%%%%%

In this section we have thoroughly tested the capabilities of our CFD code in predicting the fluid forces in the case of two coaxial cylinders. In what follows we challenge the code with a more complicated configuration consisting of a tube bundle immersed in a viscous fluid.

%%%%%%%%%%%%%%%%%%%%%%%%%%%%%%%%%%%%%%%%%%%%%%%%%%%%%%%%%%%%%%%%
%%%%%%%%%%%%%%%%%%%%%%%%%%%%%%%%%%%%%%%%%%%%%%%%%%%%%%%%%%%%%%%%

\section{Study 2. Vibrations of a cylinder in a square tube bundle immersed in a viscous fluid}\label{sec:4}
The importance of fluid-elastic forces in tube bundle vibrations can hardly be over-emphasized, in view of their damaging potential. In the last decades, advanced models for representing the fluid-elastic forces through added-coefficients have therefore been developed by the community of the domain. Those models are nowadays embedded in the methodologies that are used on a regular basis by both steam generators providers and operators, in order to prevent the risk of a tube failure with adequate safety margins. From an R\&D point of view however, the need still remains for more advanced models of the fluid added-coefficients, in order to fully decipher the physics underlying the observed phenomena.  

In what follows, we aim to determine numerically and experimentally (new measurement setup built at CEA) the effect of the Stokes number $Sk$ on the fluid coefficients, considering the case of a square tube bundle immersed in a viscous fluid at rest. The numerical results obtained with TrioCFD are thoroughly compared with the experimental results and also a theoretical approach proposed by~\cite{Pettigrew1995}. 

\subsection{Presentation of the problem}
\begin{figure}[!ht]
\begin{centering}
\begin{tabular}{cc}
\includegraphics[width=0.45\linewidth]{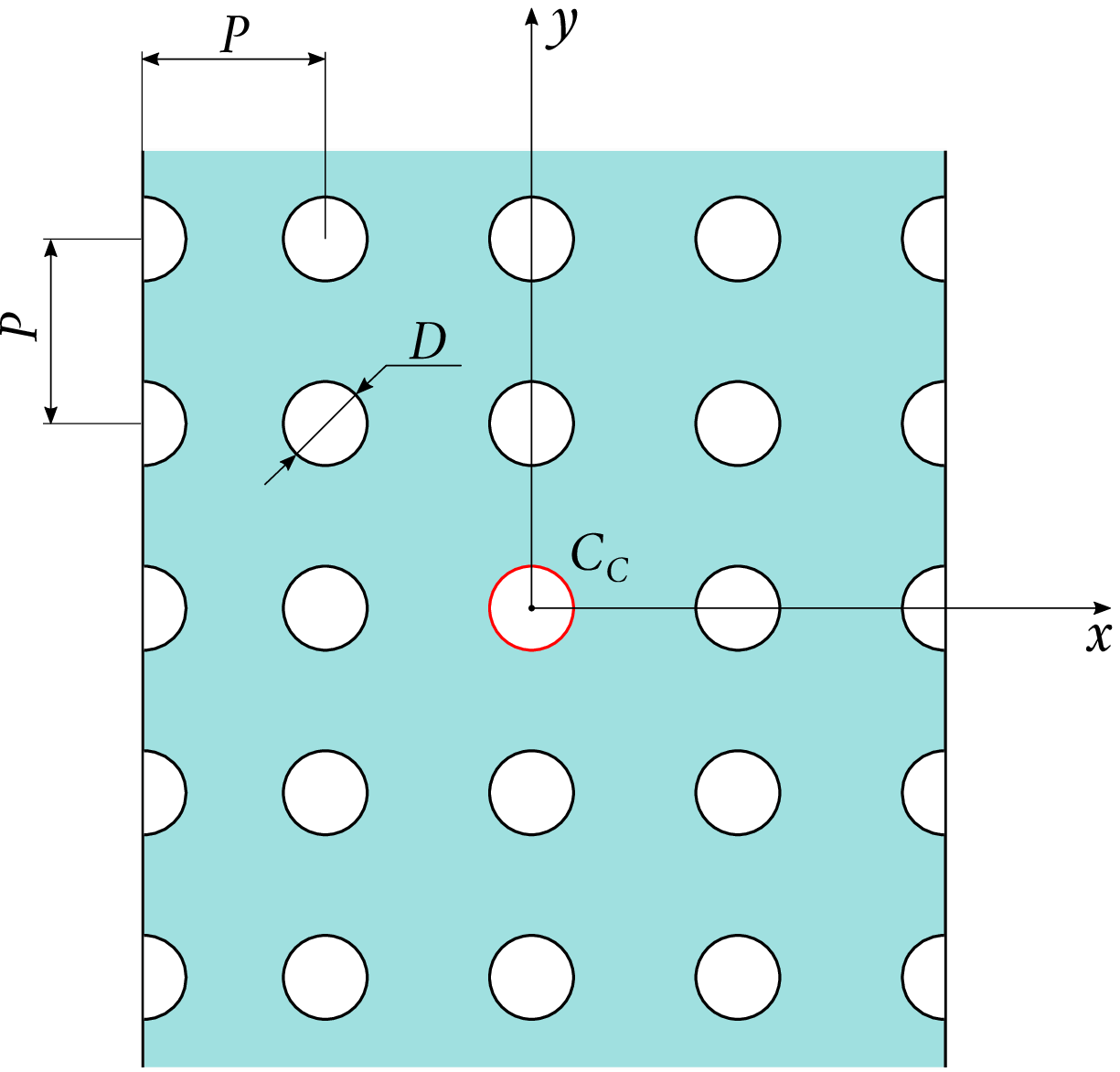} & \includegraphics[width=0.49\linewidth]{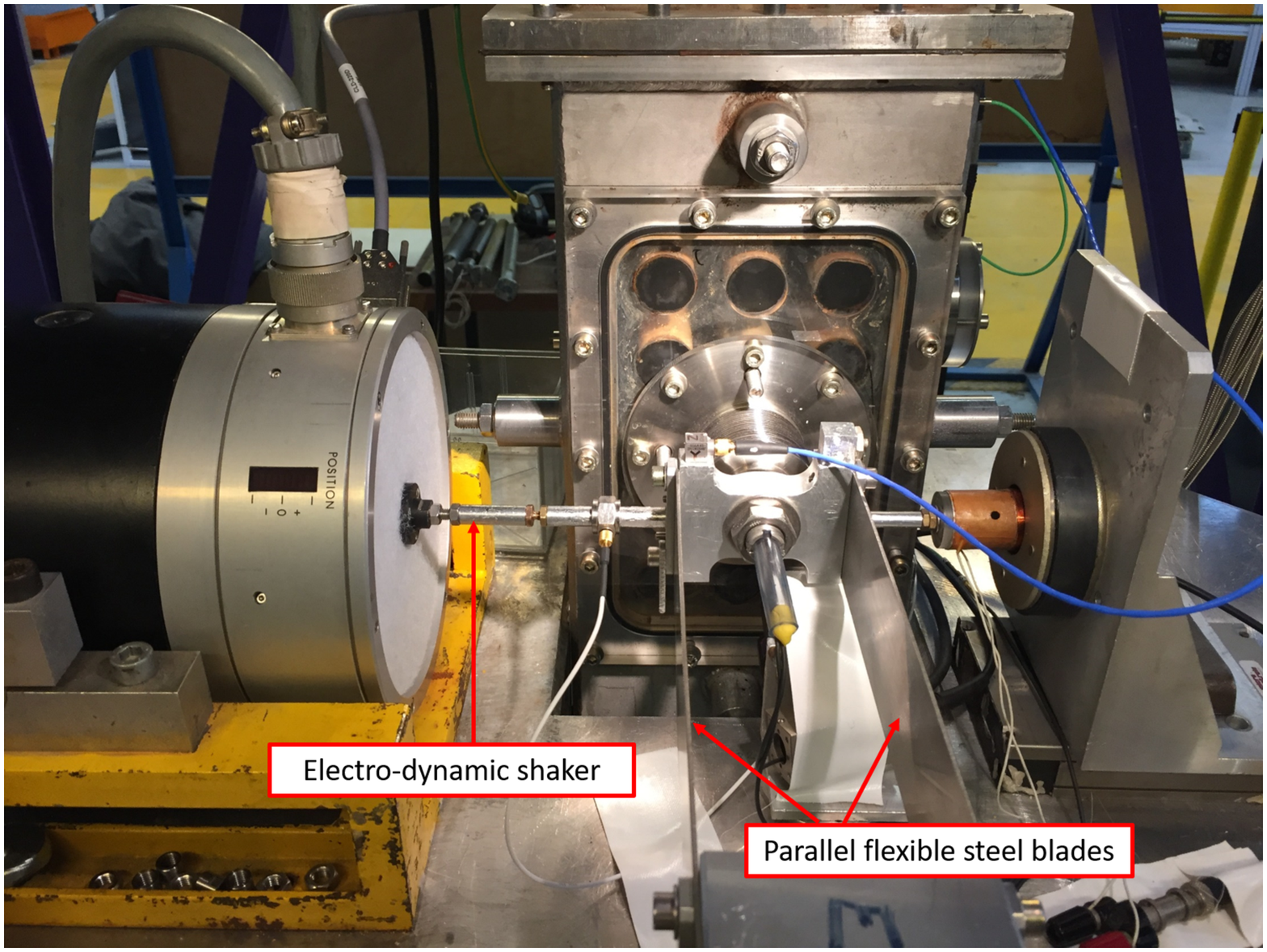}
\tabularnewline
(a) & (b)
\tabularnewline
\end{tabular}
\par\end{centering}
\protect\caption{Study 2. (a) Configuration with $3 \times 5$ cylinders plus two columns of 5 half-cylinders at the boundaries. The tube bundle is immersed in a homogeneous fluid with mass density $\rho$ and kinematic viscosity $\nu$. (b) Photo of the experimental setup designed to measure the fluid coefficients by the direct method.}
\label{fig:Initial_problem_2}
\end{figure}

Let $C_C$ be a cylinder of diameter $D$, located in the central position of a square tube bundle with pitch $P$, see Fig.~\ref{fig:Initial_problem_2} a). The cylinder $C_C$ oscillates in the $\left(x,y\right)$ plane with a simple harmonic motion of angular frequency $\Omega$ and a displacement amplitude $U$. All the other cylinders are stationary. The tube bundle is immersed in a Newtonian and homogeneous fluid, with mass density $\rho$ and kinematic viscosity $\nu$. The fluid flow generated by the oscillation of the central cylinder is assumed as incompressible and two-dimensional.
 
Following the approach presented in \S\ref{sec:3} with $D_1=D_2=D$, the dimensionless fluid force acting on the central cylinder writes
\begin{equation}
\textbf{f}^*\\
= -\pi\left(m_{self}
\frac{d^2{\bf{u}}^*}{dt^{*2}}
 + c_{self} \frac{d{\bf{u}}^*}{dt^{*}}
\right), 
\end{equation}
with $m_{self}$ and $c_{self}$ the self-added mass and damping dimensionless coefficients, respectively. For a given configuration, these coefficients are functions of the pitch ratio $P/D$, the Keulegan-Carpenter number $KC=U/D$ and the Stokes number $Sk=D^2\Omega/(2\pi/\nu)$.\\

In order to determine the variations of $m_{self}$ and $c_{self}$ with $Sk$, an experimental setup consisting of $3\times 5$ cylinders (plus two columns of 5 half-cylinders at the boundaries) has been built at CEA, see Fig.~\ref{fig:Initial_problem_2}(b). This setup is briefly presented in the following and the reader is referred to~\cite{Caillaud1999,Caillaud2003,Piteau2012} for an extensive description. Following the principle of the direct method developed by~\cite{Tanaka1981}, an electro-dynamic shaker (PRODERA 200 N) is used to impose a harmonic motion to the central tube, supported by two parallel flexible steel blades allowing large vibrations in the $x$ direction. The fluid coefficients are directly extracted from the measure of the fluid force, obtained with two piezo-electric transducers (KISTLER 9132 A). Despite its apparent simplicity, the method of~\cite{Tanaka1981} has been progressively abandoned due to its difficulty to implement experimentally.
Still, here we reconsider this method, performing some new experiments with a tube bundle with pitch $P=0.045$ m, made with cylinders of diameter $D=0.03$ m and length $L=0.3$ m. The central cylinder is imposed a harmonic displacement in the $x$-direction with amplitude $U=0.003$ m and a frequency $F\in\{5,10,15,20,25\}$ Hz. In terms of dimensionless numbers, this corresponds to a pitch ratio $P/D=1.5$, a Keulegan-Carpenter number $KC=10^{-1}$ and a Stokes number $Sk\in \{0.45,0.90,1.35,1.80,2.25\}\times 10^4$.\\ 

In what follows, we aim to compare the experimental results with TrioCFD predictions. In the discussion of our results we also refer to the phenomenological estimations of the fluid coefficients provided by~\cite{Pettigrew1995} for $KC<<1$
\begin{equation}\label{eq:theoretical_solution_2}
    m_{self} = \dfrac{\pi}{4}\dfrac{(D_e/D)^2+1}{(D_e/D)^2-1}, \quad c_{self} = \dfrac{4}{\sqrt{\pi}}\dfrac{1}{\sqrt{Sk}}\dfrac{1+(D/D_e)^3}{\left[1-(D/D_e)^2\right]^2},
\end{equation}
where $D_e= P\left(1.07 + 0.56 P/D\right)$ represents an equivalent diameter introduced to model the confinement of the tube bundle.

\subsection{Numerical setup}
The fluid domain is discretized with an unstructured grid of triangles generated by the SALOMÉ platform, see \cite{Ribes2007}.

Before analyzing the meshes used in the numerical simulations and their properties, we start discussing on the technique used to design the geometry. In the experimental setup, the geometry is perfectly symmetrical so that there is no fluid force acting on the cylinders as they are stationary. Numerically, to ensure a zero hydrostatic force, the same discretization scheme of the fluid domain close to every single cylinder is required, while maintaining an $x$ and $y$ symmetry of the mesh grid. 
To do so, we introduce some artificial lines to subdivide the fluid domain into blocks, each of them being divided into four slices. The automatic and optimized mesh algorithm ensures that the arrangement of the generated triangles, on each side of the artificial lines, is perfectly symmetrical, so that all the slices are discretized equally.
The MG-CADSurf Module of SALOMÉ is used for this purpose. We define two mesh sizes: a local size, $lc_{fine}$, for elements close to the cylinders and a user size, $lc$ for elements far from the cylinders. A gradation parameter is also set, $\Delta = 1.1$, as the maximum ratio between the lengths of two adjacent edges.

Firstly, a mesh sensitivity analysis is carried out on a truncated fluid domain (with the last rows of cylinders cut in half) in order to reduce the time of calculation and speed up the convergence of the grid. This analysis shows that a mesh grid with $lc=0.0005$ and $lc_ {fine}=0.00015$ is suitable for an accurate estimation of the fluid forces and the related added coefficients, see \ref{sec:effect_mesh_size}. 
Then, the mesh sizes are kept constant whereas the size of the fluid domain is gradually increased by a distance of $N_D$ diameters in the $y$-directions, with $N_D$ from 1 to 5.

\subsection{Results and discussion}
The central cylinder is allowed to move in the $x$-direction with a sinusoidal displacement of five periods of time, i.e. $\mathbf{u}^* = \sin(t^*)\mathbf{e}_x$ with $t^* \in \{0, 10\pi\}$. The numerical simulations are performed for $Sk \in \{0.45,0.90,1.35,1.80,2.25\}\times 10^4$ corresponding to imposed frequencies in the range $F\in\{5,10,15,20,25\}$ Hz. The other parameters are $P/D=1.5$ and $KC = 10^{-1}$.\\

\begin{figure}[H]
	\centering
	\includegraphics[width = 0.9\linewidth]{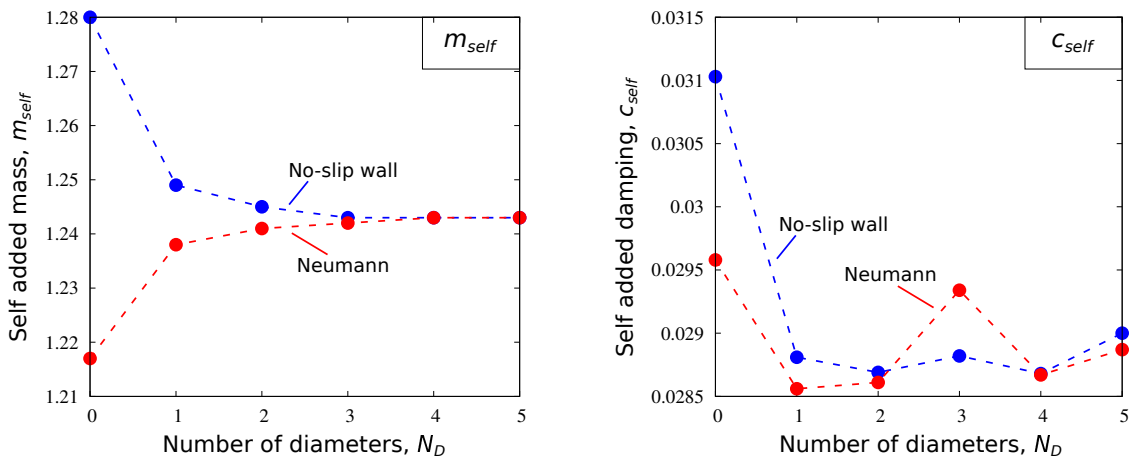}
	\caption{Study 2. Evolution of the self-added coefficients with the size of the fluid domain (the size of the fluid domain increases by a distance of $N_D$ diameters in the $y$-directions). The blue (resp. red) dashed line corresponds to the numerical predictions obtained with a wall boundary condition (resp. a Neumann boundary condition) imposed on the top and bottom boundaries of the computational domain, see Fig.~\ref{fig:Initial_problem_2}. The dimensionless parameters are $Sk = 1.8\times10^{4}$, $P/D=1.5$ and $KC = 10^{-1}$.}
	\label{fig:coefficients_Nd_BC}
\end{figure}

Before analyzing the variations of the fluid coefficients, we start discussing on the choice of the boundary conditions used in the numerical simulations. In our experiments, the fluid is laterally confined by two parallel plates. Numerically, a classical wall condition ($\mathbf{v} = \mathbf{0}$) is thus applied on the two lateral sides. The choice of the boundary condition at the top and bottom sides of the computational domain ($y$ directions) is more delicate as the tube bundle is experimentally surrounded by two columns of water. To overcome this problem, we determine the size of the computational domain which yields numerical results poorly sensitive to this choice. To do so, we test the effect of two types of $y$ boundary conditions: a wall condition and a Neumann zero stress condition ($p\mathbf{n} - \rho\nu[\nabla\mathbf{v} + (\nabla \mathbf{v})^{\operatorname{T}}]\cdot\mathbf{n}=\mathbf{0}$).  
In Fig.~\ref{fig:coefficients_Nd_BC}, we show the results of this test for $Sk = 1.8\times10^{4}$ as the size of the fluid domain increases by a distance of $N_D$ diameters in the $y$-directions. As $N_D$ increases, we observe that $m_{self}$ converges to a constant value that does not depend on the choice of the $y$ boundary condition. Similarly, the two types of boundary conditions yield very tiny differences on the values of $c_{self}$. From these observations, we thus decide to perform all our numerical simulations with $N_D = 4$ and a wall boundary condition in the $y$-directions.\\
%%%%%%%%%%%%%%%%%%%%%%%%%%%%%%%%%%%%%%%%%%%%%%%%%%%%%%%%%%
% Case 1. Evolution of added coefficients with N_D
%%%%%%%%%%%%%%%%%%%%%%%%%%%%%%%%%%%%%%%%%%%%%%%%%%%%%%%%%%

\begin{figure}[!t]
	\centering
	\includegraphics[width = 1\linewidth]{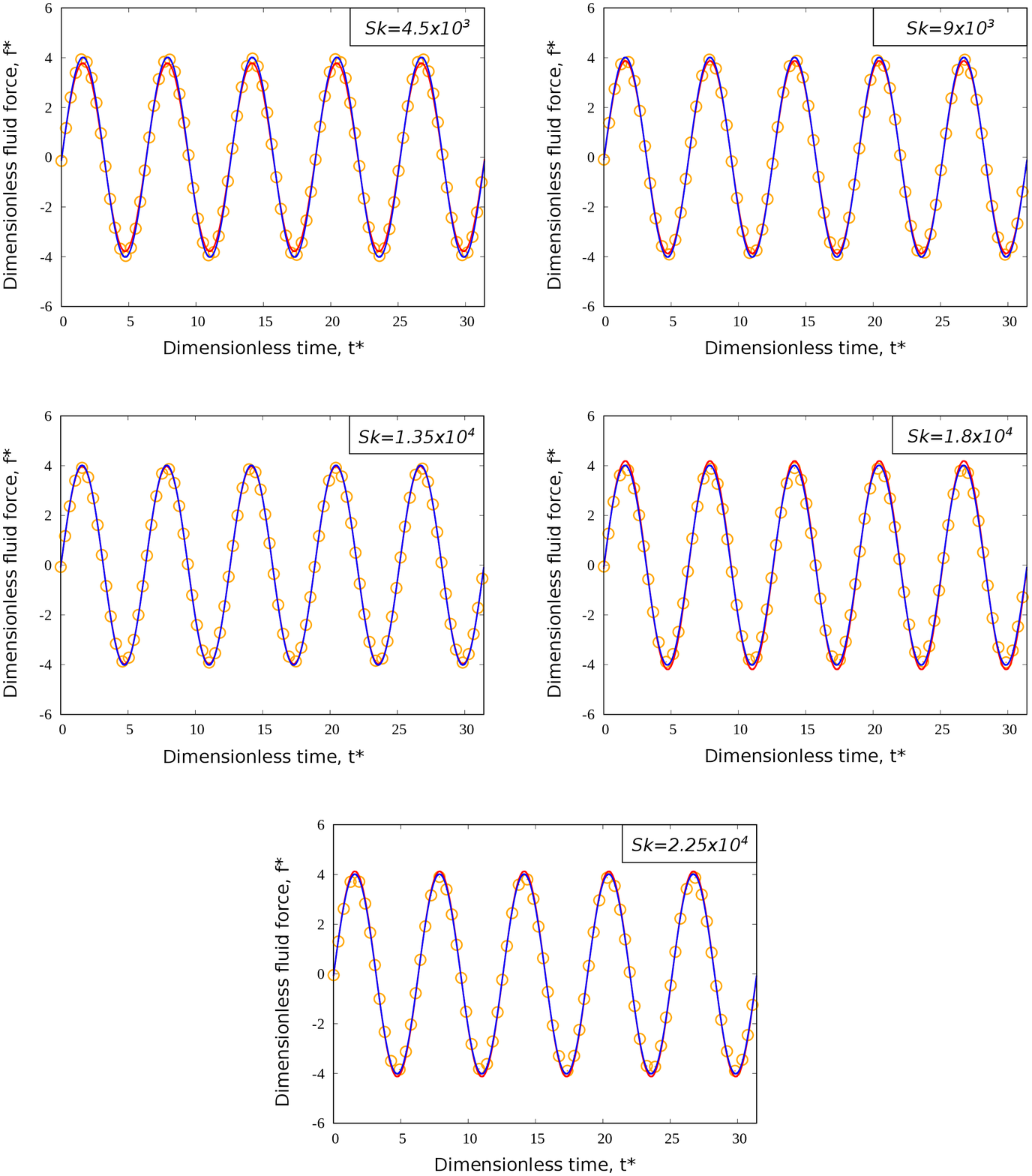}
	\caption{Study 2. Time evolution of the dimensionless fluid forces, for $Sk \in \{0.45,0.90,1.35,1.80,2.25\}\times 10^4$, $P/D = 1.5$ and $KC = 10^{-1}$. The blue solid lines correspond to the theoretical estimations of~\cite{Pettigrew1995}, see Eq.~\eqref{eq:theoretical_solution_2}. The red solid lines correspond to the experimental measurements. The orange symbols correspond to the numerical predictions. (For interpretation of the colors in this figure, the reader is referred to the web version of this article).}
	\label{fig:dimensionless_force_Sk_2}
\end{figure}

The time evolution of the dimensionless fluid force is represented in Fig.~\ref{fig:dimensionless_force_Sk_2}. We note that our simulations are in good agreement with both the theoretical and the experimental estimations, despite a small amplitude deviation, in particular with the experimental measurements for $F\in\{5,20\}$ Hz, i.e.  $Sk\in\{0.45,1.8\}\times10^4$. This small deviation will be analyzed more precisely in what follows, considering the variations of the fluid coefficients $m_{self}$ and $c_{self}$. \\ 
%%%%%%%%%%%%%%%%%%%%%%%%%%%%%%%%%%%%%%%%%%%%%%%%%%%%%%%%%%
% Evolution of forces with Sk
%%%%%%%%%%%%%%%%%%%%%%%%%%%%%%%%%%%%%%%%%%%%%%%%%%%%%%%%%%

\begin{figure}[!t]
	\centering
	\includegraphics[width = 0.9\linewidth]{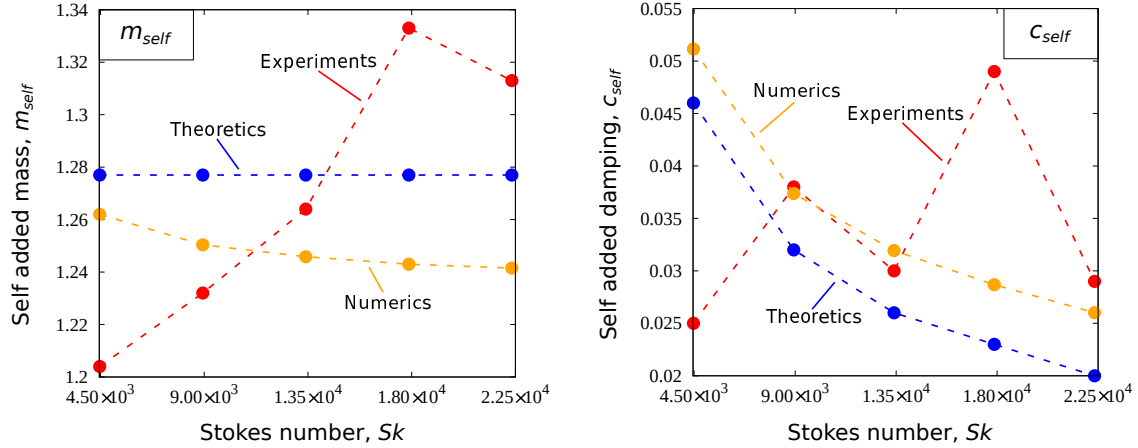}
	\caption{Study 2. Evolution of the self-added coefficients with the Stokes number $Sk$, for $P/D = 1.5$ and $KC = 10^{-1}$. The red, blue and orange dashed lines with points refer respectively to the experimental measurements, the theoretical estimations of~\cite{Pettigrew1995} and the numerical predictions.}
	\label{fig:coefficients_Sk_2}
\end{figure}

The evolution of the self-added coefficients with the Stokes number, $Sk$, is depicted in Fig.~\ref{fig:coefficients_Sk_2}. Our simulations are in good agreement with the theoretical estimations, in the sense that similar trends are recovered, bringing out the same behavior of the fluid coefficients. Still, we note that the simulations tend to underestimate (resp. overestimate) $m_{self}$ (resp. $c_{self}$), with a relative deviation always smaller than $\iota \le 3.1\%$ (resp. $\iota \le 27.5\%)$, see Tab.~\ref{tab:Comparison_theory_numerics_Sk_2}. We attribute these deviations to the fact that the theoretical approach of~\cite{Pettigrew1995} is based on a strong approximation in which the tube bundle is modeled as a system of two concentric cylinders, through the definition of a fictive equivalent diameter determined empirically. Also, the theory of~\cite{Pettigrew1995} relies on a linear approach, contrarily to our numerical simulations in which the nonlinear convective term of the dimensionless Navier-Stokes equation is taken into account. 
The agreement between the numerical predictions and our experimental results is partial, showing a good agreement for the self-added mass coefficient $m_{self}$, with $\iota<7\%$, but some important deviation for $c_{self}$, especially for the lowest Stokes number $Sk=0.45\times10^4$, where $\iota\approx105\%$. It shall be noted that it is very delicate to measure the self-damping coefficient in our experiments as the determination of $c_{self}$ relies on a tiny phase difference between the fluid force and the imposed displacement. Also, for such a low Stokes number, i.e. low frequency, our experiments become very sensitive to parasitic frequencies in the experimental setup, leading to a bad signal to noise ratio. Still, these first comparisons are very encouraging and should foster further developments of the experimental direct method.\\
%%%%%%%%%%%%%%%%%%%%%%%%%%%%%%%%%%%%%%%%%%%%%%%%%%%%%%%%%%
% Evolution of added coefficients with Sk
%%%%%%%%%%%%%%%%%%%%%%%%%%%%%%%%%%%%%%%%%%%%%%%%%%%%%%%%%%
%%%%%%%%%%%%%%%%%%%%%%%%%%%%%%%%%%%%%%%%%%%%%%%%%%%%%%%%%%%%%%%%%%%%%%%%%%%%%%%%%%%%%%%%%%%%%%%%%%%
%%%%%%%%%%%%%%%%%%%%%%%%%%%%%%%%%%%%%%%%%%%%%%%%%%%%%%%%%%
% Table of comparison of the added coefficients with Sk
%%%%%%%%%%%%%%%%%%%%%%%%%%%%%%%%%%%%%%%%%%%%%%%%%%%%%%%%%%
\begin{table}[!t]
\begin{center}
    \includegraphics[width=0.4\linewidth]{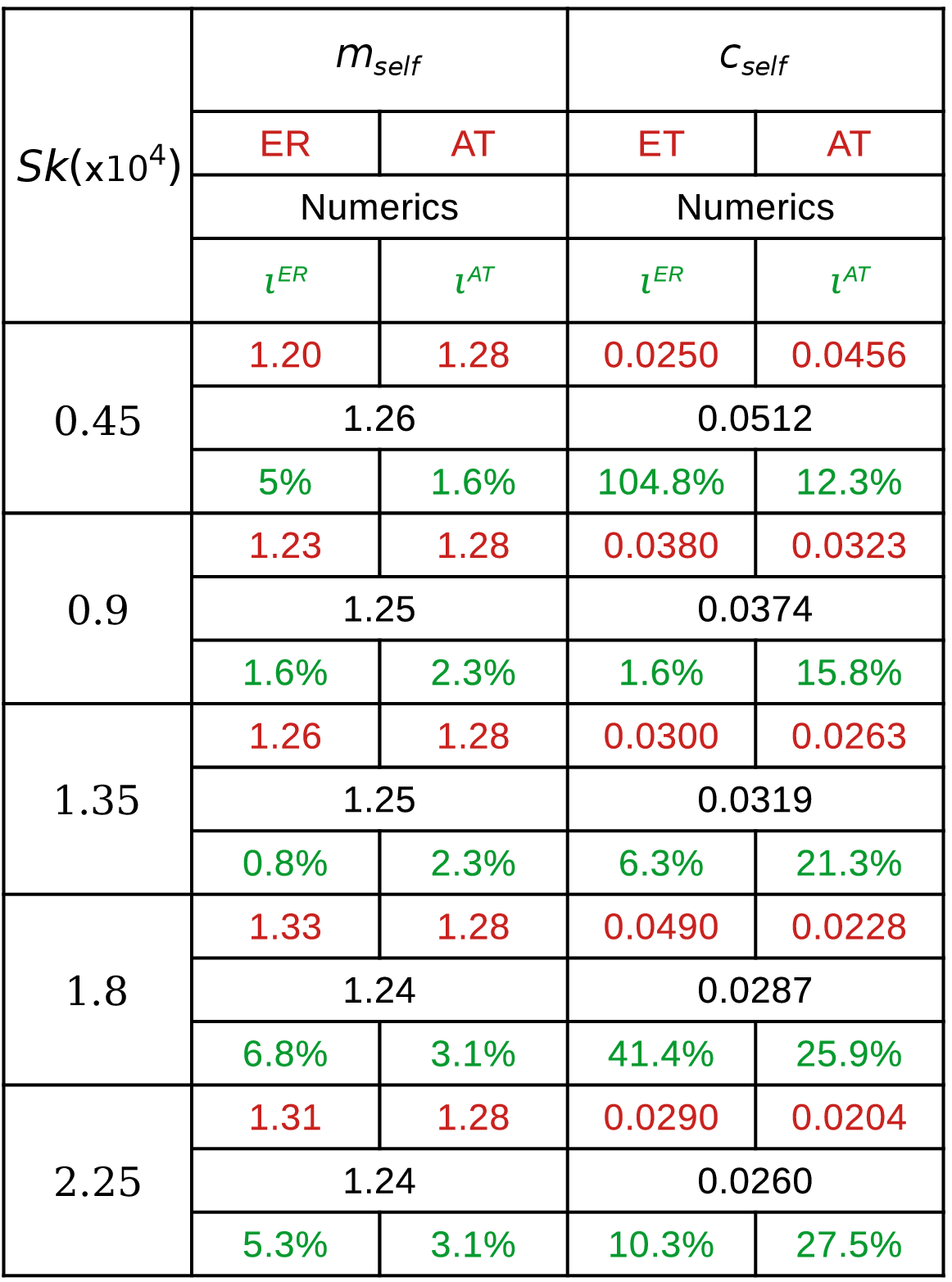}
    \caption{Study 2. Table of the self-added coefficients as a function of the Stokes number, $Sk$ and relative deviation, $\iota$. The notations ER and AT refer to the experimental results and the approximate theory. The pitch ratio is $P/D = 1.5$ and the Keulegan-Carpenter number is $KC = 10^{-1}$.}\label{tab:Comparison_theory_numerics_Sk_2}
\end{center}
\end{table}

\begin{figure}[!t]
\centering
	\includegraphics[scale=0.3]{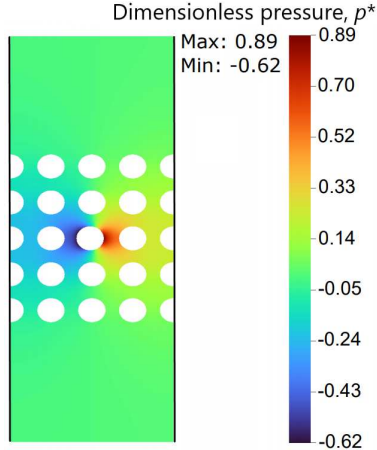}
	\caption{Study 2. Dimensionless pressure field, for $Sk = 1.8\times10^{4}$, $P/D = 1.5$ and $KC = 10^{-1}$, computed with TrioCFD.}
	\label{fig:pressure_field_DIVA}
\end{figure}

As a conclusion to study 2, we show in Fig. \ref{fig:pressure_field_DIVA} the pressure distribution when $C_C$ is imposed a sinusoidal displacement in the $x$-direction, for $Sk = 1.8\times10^{4}$, $P/D = 1.5$ and $KC = 10^{-1}$. Note that this snapshot is taken at a time when $C_C$ has a positive acceleration. As expected from the inertia effect of the added mass, the fluid  pushes $C_C$ in a direction opposite to its acceleration vector, by creating a negative (resp. positive) pressure on the left (resp. right) side of $C_C$. 

%%%%%%%%%%%%%%%%%%%%%%%%%%%%%%%%%%%%%%%%%%%%%%%%%%%%%%%%%%%%%%%%%%%%%%%%%%%%%%%%%%%%%%%%%%%%%%%%%%%
%%%%%%%%%%%%%%%%%%%%%%%%%%%%%%%%%%%%%%%%%%%%%%%%%%%%%%%%%%%%%%%%%%

\section{Conclusion and perspectives}\label{sec:5}
In this study, we have first considered the vibration of two coaxial cylinders separated by a viscous fluid. The ALE numerical approach, based on a mesh update technique that considers the motion of the boundary, has been carried out with TrioCFD to estimate the fluid forces acting on the two cylinders. 

We have started studying the case in which the outer cylinder is stationary while the inner one is imposed a harmonic motion. The numerical predictions for the fluid forces and the corresponding added coefficients are in very good agreement with the theoretical estimations, even if a tiny difference is observed in the range of low Stokes numbers in which the theoretical estimations reach their limit of validity.
More specifically, we have shown that the two forces are in phase opposition and sensitive to the Stokes number, $Sk$, and the dimensionless separation distance, $\varepsilon$. The amplitude (resp. phase) of the forces decreases (resp. increases) as $Sk$ or $\varepsilon$ increases, recovering the inviscid limit for large Stokes numbers or the limit of an isolated cylinder for large $\varepsilon$. The fluid coefficients variations are also correctly reproduced by the numerical simulations, which are shown to respect the scale invariance expected from the dimensional analysis. 
Secondly, we have studied the case in which the inner cylinder is stationary while the outer one is imposed a harmonic motion. From this study, we have shown that the force acting on the stationary cylinder does not depend on which cylinder is moving (inner or outer).

After having successfully tested the capabilities of the ALE method implemented in TrioCFD, in the configuration of two coaxial cylinders, we have considered the problem of a vibrating cylinder located in the central position of a square tube bundle immersed in a viscous fluid. We have shown that the numerical predictions for the self-added mass coefficient are in good agreement with the results of our experiments led at CEA and a theoretical estimation used by engineers. The numerical predictions for the self-added damping coefficient are also in good agreement with the theoretical estimation. They partially agree with the experimental results, showing an important divergence for low Stokes numbers. However, such a deviation is mainly due to our experimental setup, which exhibits a strong sensitivity to parasitic frequencies for low Stokes numbers.  

In conclusion, the ALE method implemented in TrioCFD is particularly efficient in solving fluid-structure interaction problems with a fluid initially at rest. We are now confident that further developments will make it possible to tackle three-dimensional problems with an incident fluid flow.

%%%%%%%%%%%%%%%%%%%%%%%%%%%%%%%%%%%%%%%%%%%%%%%%%%%%%%%%%%%%%%%%%%%%%%%%%%%%%%%%%%%%%%%%%%%%%%%%%%%%%%%%%%%%%%%%%%%%%%%%%%%%%%%%%%%%
%%%%%%%%%%%%%%%%%%%%%%%%%%%%%%%%%%%%%%%%%%%%%%%%%%%%%%%%%%%%%%%%%%%%%%%%%%%%%%%%%%%%%%%%%%%%%%%%%%%%%%%%%%%%%%%%%%%%%%%%%%%%%%%%%%%%
\section*{Acknowledgements}
The authors fully acknowledge P. Piteau and T. Valin (CEA Saclay) for implementing the Tanaka's direct method of measurement on the DIVA experimental setup located at CEA Saclay, and for sharing their results as part of the present study. The authors also acknowledge financial support for this work,
which was performed in the framework of a joint research
program co-funded by FRAMATOME, EDF and CEA (France). 

\section*{Declaration of competing interest}
The authors declare that they have no known competing financial interests or personal relationships that could have appeared to influence the work reported in this paper.

%%%%%%%%%%%%%%%%%%%%%%%%%%%%%%%%%%%%%%%%%%%%%%%%%%%%%%%%%%%%%%%%%%%%%%%%%%%%%%%%%%%%%%%%%%%%%%%%%%%%%%%%%%%%%%%%%%%%%%%%%%%%%%%%%%%%
%%%%%%%%%%%%%%%%%%%%%%%%%%%%%%%%%%%%%%%%%%%%%%%%%%%%%%%%%%%%%%%%%%%%%%%%%%%%%%%%%%%%%%%%%%%%%%%%%%%%%%%%%%%%%%%%%%%%%%%%%%%%%%%%%%%%
\appendix
%%%%%%%%%%%%%%%%%%%%%%%%%%%%%%%%
\section{Discretization of the non-linear convection term: MUSCL scheme} \label{sec:muscl}
\begin{figure}[h!]
\centering
\begin{tikzpicture}[scale=0.6]%,cap=round,>=latex]

\coordinate  (S1) at (3cm,0cm);
\coordinate  (S2) at (6cm,5cm);
\coordinate  (A) at (2cm,6cm);
\coordinate  (B) at (8cm,2cm);
\coordinate  (Cj) at (3.66cm,3.66cm);
\coordinate  (Ci) at (5.66cm,2.33cm);
\coordinate  (xj) at (2.5cm,3cm);
\coordinate  (xi) at (4.5cm,2.5cm);
\coordinate  (M) at (3.33cm,1.83cm);

\draw (S1) -- (S2) -- (A) --  (S1);
\draw (S1) -- (S2) -- (B) --  (S1);
\draw [dashed] (S1) -- (Cj) -- (S2) --  (Ci) -- (S1);

\draw [thick, red] (Cj) -- (S1) node[near start, right] {$\gamma_{ij}$};

\draw (Ci) node[below] {$G_i$};
\draw (Cj) node[above] {$G_j$};
\draw (Ci) node {$\bullet$};
\draw (Cj) node {$\bullet$};

\draw (xi) node[right] {$x_i$};
\draw (xj) node[left] {$x_j$};
\draw (xi) node {$\bullet$};
\draw (xj) node {$\bullet$};

\draw (S1) node[left] {$S_1$};
\draw (S2) node[right] {$S_2$};
\draw (S1) node {$\bullet$};
\draw (S2) node {$\bullet$};

\draw (M) node[left] {$M_{ij}$};
\draw (M) node {$\bullet$};

\draw (6.5,2.2) node[right] {$K_i$};
\draw (2.2,5.3) node[right] {$K_j$};
\draw (5.4,3.6) node[right] {$w_i$};
\end{tikzpicture}
\caption{Control volume $w_i$ associate to the DoF $x_i$}
\label{vdc}
\end{figure}
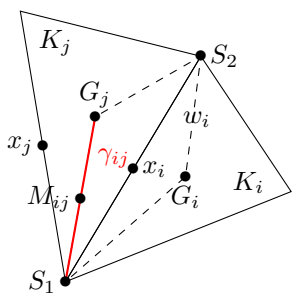

In this paragraph, in order to simplify the presentation, we restrict ourselves to the two-dimensional case. We recall that in the VDF formulation with each degrees of freedom $ x_i $ of the speed, we associate a control volume $w_i$ (see Fig.~\ref{vdc}). We have:
$$\int_{w_i} \nabla \cdot (\mathbf{v} \otimes \mathbf{v}) dV =  \int_{\gamma_i} \mathbf{v} (\mathbf{v} \cdot \mathbf{n} ) d\sigma,  $$
where $\gamma_i$ represents the faces of the control volume $w_i$.

In the MUSCL scheme (Monotone Upstream-Centred Scheme for Convective flows)~\cite{van1979towards}, the flow across the face $\gamma_{ij}$ is approximated based on the Simpson interpolation formula:
$$\int_{\gamma_{ij}} \mathbf{v} (\mathbf{v} \cdot \mathbf{n} ) d\sigma \approx \frac{ \mid  \gamma_{ij} \mid}{6}  \left( \mathbf{v}_{S_1} \mathbf{V}_{S_1}  + 4 \mathbf{v}_{M_{ij}} \mathbf{V}_{M_{ij}} +   \mathbf{v}_{G_j} \mathbf{V}_{G_j} \right), $$
where we have used the notations shown in Fig. $\ref{vdc}$ and, if $\mathbf{v}_{M_{ij}} \cdot \mathbf{n}_{ij} > 0$:
\begin{equation*}
\left\{
\begin{aligned}
\mathbf{V}_{S_1} &=  \mathbf{v}_{i} + \mid x_j S_1 \mid  \nabla\mathbf{v}_i,\\
\mathbf{V}_{M_{ij}} &=  \mathbf{v}_{i} + \mid x_j M_{ij} \mid \nabla\mathbf{v}_{i} , \\
\mathbf{V}_{G_j} &= 2 \mathbf{v}_{M_{ij}} - \mathbf{v}_{S_1} ,
\end{aligned} 
\right.
\end{equation*}
else, 
\begin{equation*}
\left\{
\begin{aligned}
\mathbf{V}_{S_1} &=  \mathbf{v}_{j} + \mid x_j S_1 \mid \nabla\mathbf{v}_j,\\
\mathbf{V}_{M_{ij}} &=  \mathbf{v}_{j} + \mid x_j M_{ij} \mid \nabla\mathbf{v}_{j} , \\
\mathbf{V}_{G_j} &= 2 \mathbf{v}_{M_{ij}} - 2\mathbf{v}_{S_1}.
\end{aligned}
\right.
\end{equation*}  

Different slope limiters are used in order to calculate the  $\nabla\mathbf{v}_i$ from the gradients associated with the triangles $K_i$ and $K_j$ sharing the face $S_1S_2$:
\begin{equation*}
\nabla\mathbf{v}_{i} = \begin{cases}  
    \text{minmod}(\nabla\mathbf{v}_{K_i}, \nabla\mathbf{v}_{K_j}) ,\\
    \text{Van-Leer} (\nabla\mathbf{v}_{K_i}, \nabla\mathbf{v}_{K_j}) , \\
    \text{Van-Albanda} (\nabla\mathbf{v}_{K_i}, \nabla\mathbf{v}_{K_j}). \end{cases}
\end{equation*} 
with:
\begin{equation*}
\begin{aligned}
\text{minmod}(a,b) &:=\begin{cases}0 & \text{if } a \cdot b \leq 0, \\ 
a & \text{if } \mid a \mid <   \mid b \mid , \quad a \cdot b > 0,\\
b &  \text{if } \mid a \mid >  \mid b \mid , \quad a \cdot b > 0.\end{cases}\\\\
\text{Van-Leer}(a,b) &:=\begin{cases}0 & \text{if } a \cdot b \leq 0, \\
\dfrac{2 a \cdot b}{ a + b} & \text{otherwise}.\end{cases}\\\\
\text{Van-Albanda}(a,b) &:=\begin{cases}0 & \text{if } a \cdot b \leq 0, \\ 
\dfrac{ a \cdot b \cdot (a+b)}{ a^2 + b^2} & \text{otherwise}.\end{cases}\\
\end{aligned}
\end{equation*}

\section{Calculation of the time step}\label{sec:D}
Flexibility of the code lets us to choose the most appropriate time integration schemes, with implicit or explicit temporal time marching. 

The explicit schemes require an additional stability criteria (Courant-Friedrichs-Lewy condition) over the time step of the simulation, $\Delta t$: 
\begin{equation}
    \Delta t < CFL \times \Delta t_{stab},
\end{equation}
where $CFL=\mathcal{O}(1)$ and the time step $\Delta t_{stab}$ is calculated as the harmonic mean of the convection time step $\Delta t_{conv}$ and diffusion one $\Delta t_{diff}$ as: 
\begin{equation}
    \dfrac{1}{\Delta t_{stab}} =  \dfrac{1}{\Delta t_{conv}} + \dfrac{1}{\Delta t_{diff}},
\end{equation}
with:
\begin{equation}
    \left \{
    \begin{aligned}
    \Delta t_{conv} &= \dfrac{\Delta x}{\max(||\mathbf{v}||)}\\
    \Delta t_{diff} &= \dfrac{\Delta x^2}{2\nu}
    \end{aligned}
    \right.
\end{equation}
and $\Delta x$ the minimum mesh size of the fluid domain.

Thus, the time step is calculated as:
\begin{equation}
    \Delta t = \min (\Delta t _{stab}, \Delta t_{max}) \times CFL,   
\end{equation}
where $\Delta t_{stab}$ is calculated by TrioCFD before each time step and $\Delta t_{max}$ is set by the user.  

We can see that the restriction on the stability time step is due to the diffusion time step, especially in the case of a fine mesh (quadratic term in $\Delta x$). Thus, an implicit treatment of the diffusion term can considerably reduce the time of calculation. To overcome this problem, one can use an implicit scheme, as in our case, or can implicitly treat the diffusion term, at each time step using a conjugate gradient approach. The time step is then calculated only from the convection time step (when an implicit scheme is used, the time step is calculated as if the convection is completely explicit even if it is semi-implicit).

\section{Study 1. Further analyses} 
In this Appendix, we report further results obtained for the case study 1.\\

\begin{figure}[!ht]
	\centering
	\includegraphics[width = 0.43\linewidth]{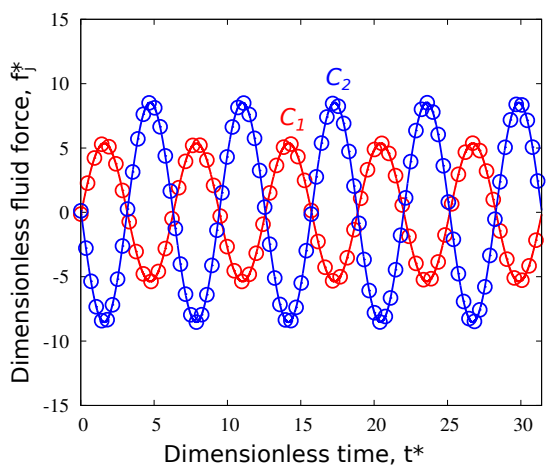}
	\caption{Study 1. Testing the scale invariance. Time evolution of the dimensionless fluid forces, for $Sk=10^4$, $\varepsilon = 2$ and $KC = 10^{-2}$. The solid lines correspond to the numerical predictions performed with $D_1 = 1$ unit of length. The symbols correspond to the numerical predictions performed with $D_1 = 0.0316$ unit of length.}
	\label{fig:dimensionless_force_inv}
\end{figure}

Firstly, we check the scale invariance of the numerical results, considering two self similar configurations in terms of dimensionless numbers: $\varepsilon = 2$, $Sk = 10^4$ and $KC = 10^{-2}$. The time evolution of the dimensionless fluid forces is represented in Fig.~\ref{fig:dimensionless_force_inv}, for both cases. As expected, the numerical predictions are strictly identical for both configurations, confirming that the dimensionless added coefficients are functions of $Sk$, $\varepsilon$ and $KC$.\\

\begin{figure}[!ht]
\begin{centering}
\begin{tabular}{c c c}
\includegraphics[width = 0.43\linewidth]{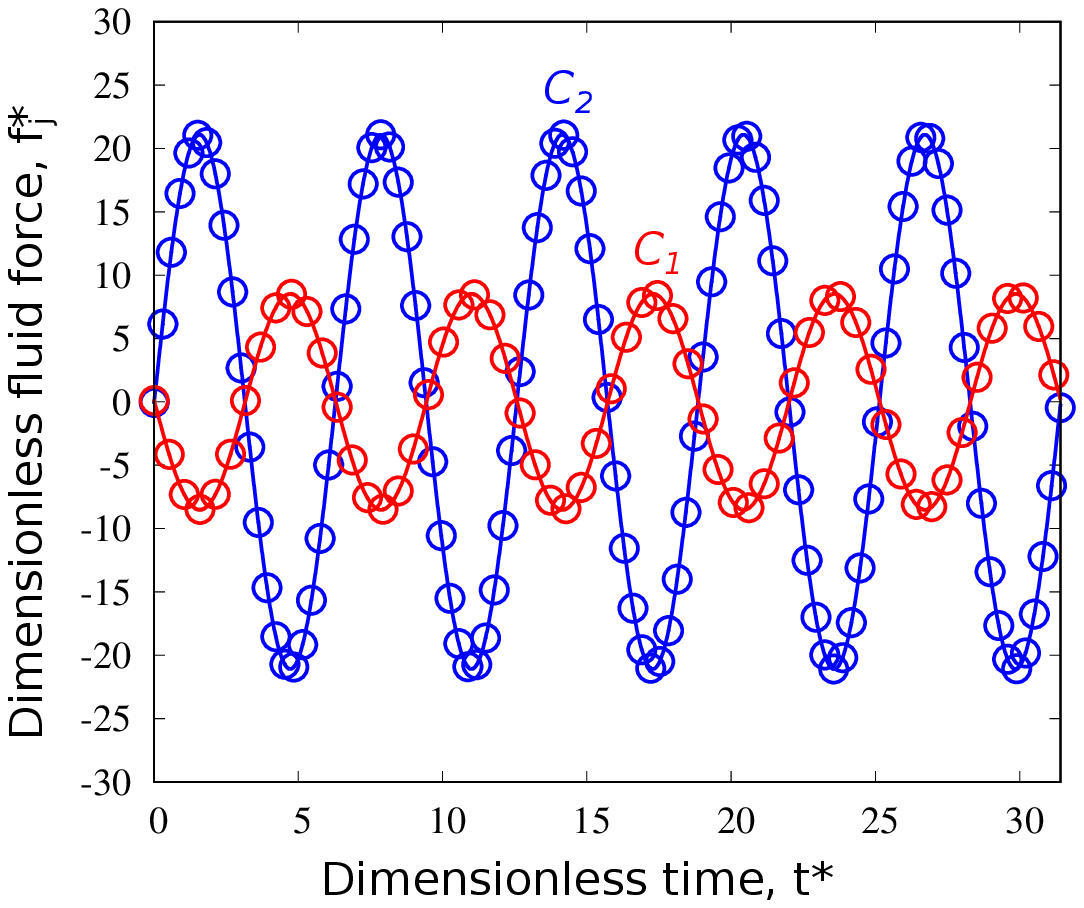} & 
&
\includegraphics[width = 0.43\linewidth]{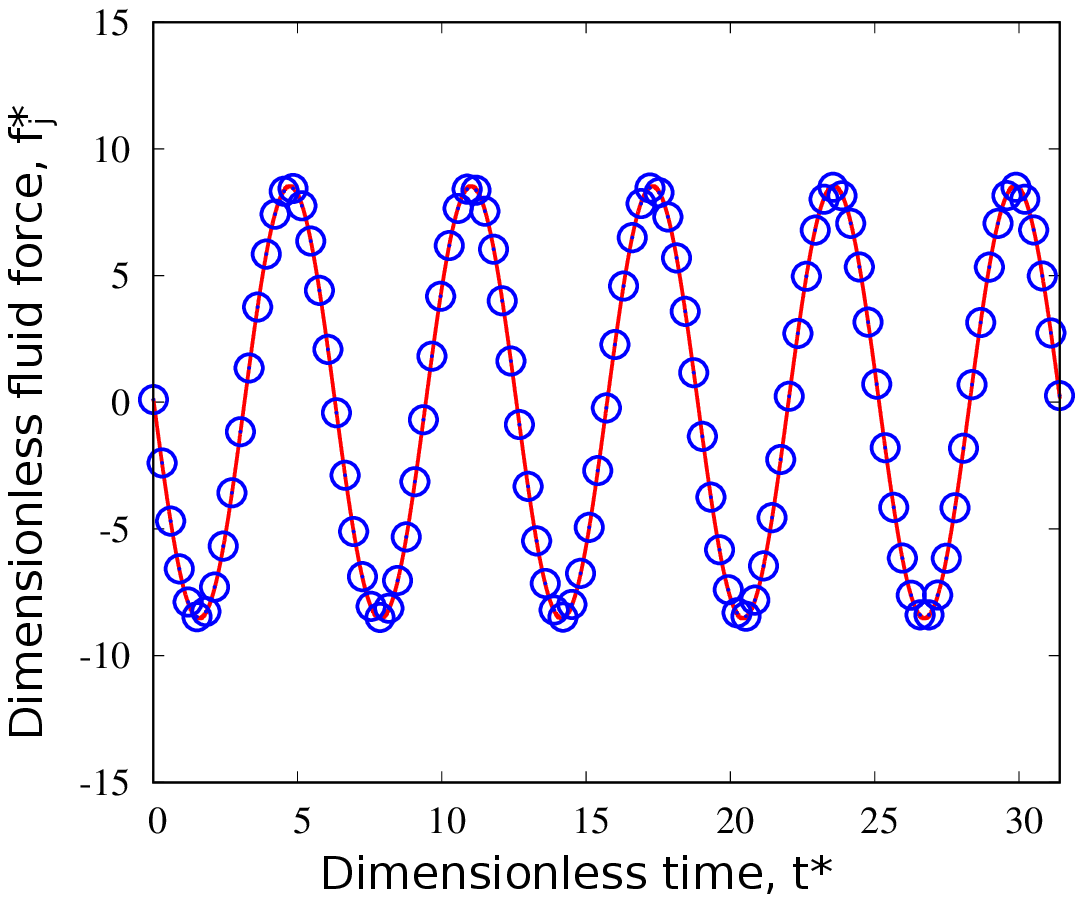}
\tabularnewline
(a) & & (b)
\tabularnewline
\end{tabular}
\par\end{centering}
\protect\caption{Study 1. Testing the symmetry of the fluid added mass and damping matrices. (a) Time evolution of the dimensionless fluid forces, for $Sk = 10^{4}$, $\varepsilon = 2$ and $KC = 10^{-2}$. The red and blue solid lines correspond to the theoretical estimations of~\cite{Yeh1978}, see Eq.~\eqref{eq:exact_theory}. The symbols correspond to the numerical predictions. (b) Time evolution of the dimensionless fluid force on the stationary cylinder, for $Sk = 10^{4}$, $\varepsilon = 2$ and $KC = 10^{-2}$. The solid line corresponds to the numerical predictions as $C_1$ oscillates. The symbols correspond to the numerical predictions as $C_2$ oscillates.}
\label{fig:dimensionless_force_outer}
\end{figure}

Then, we check the symmetry of the fluid added mass and damping matrices for $\varepsilon = 2$, $Sk = 10^4$ and $KC = 10^{-2}$, imposing a sinusoidal displacement on the outer cylinder. The time evolution of the dimensionless fluid forces is represented in Fig.~\ref{fig:dimensionless_force_outer}(a). Here again, a very good agreement between theoretical estimations and numerical predictions is observed. Also, Fig.~\ref{fig:dimensionless_force_outer}(b) shows that the dimensionless fluid force acting on $C_1$ as $C_2$ moves is the same as the dimensionless fluid force acting on $C_2$ as $C_1$ moves.

%%%%%%%%%%%%%%%%%%%%%%%%%%%%%%%%%%%%%%%%%%%%%%%%%%%%%%%%%%
% Case 1. Test of invariance.
%%%%%%%%%%%%%%%%%%%%%%%%%%%%%%%%%%%%%%%%%%%%%%%%%%%%%%%%%%

%%%%%%%%%%%%%%%%%%%%%%%%%%%%%%%%%%%%%%%%%%%%%%%%%%%%%%%%%
%Case 2. Evolution of cross forces: check symmetry of matrices
%%%%%%%%%%%%%%%%%%%%%%%%%%%%%%%%%%%%%%%%%%%%%%%%%%%%%%%%%%

\section{Study 1. Tables of comparison numerics versus theory} \label{sec:AppendixA_numerics_vs_theory}
In this Appendix, we report the tables of comparison between the theoretical and numerical values of the fluid added coefficients for the case study 1, with two coaxial vibrating cylinders. The relative deviation $\iota$ is also reported in the tables. The notations ET and AT refer to the exact and asymptotic theories.
%%%%%%%%%%%%%%%%%%%%%%%%%%%%%%%%%%%%%%%%%%%%%%%%%%%%%%%%%%
% Case 1. Evolution of added coefficients with Sk
%%%%%%%%%%%%%%%%%%%%%%%%%%%%%%%%%%%%%%%%%%%%%%%%%%%%%%%%%%
\begin{table}[!ht]
\begin{center}
    \includegraphics[width=0.675\linewidth]{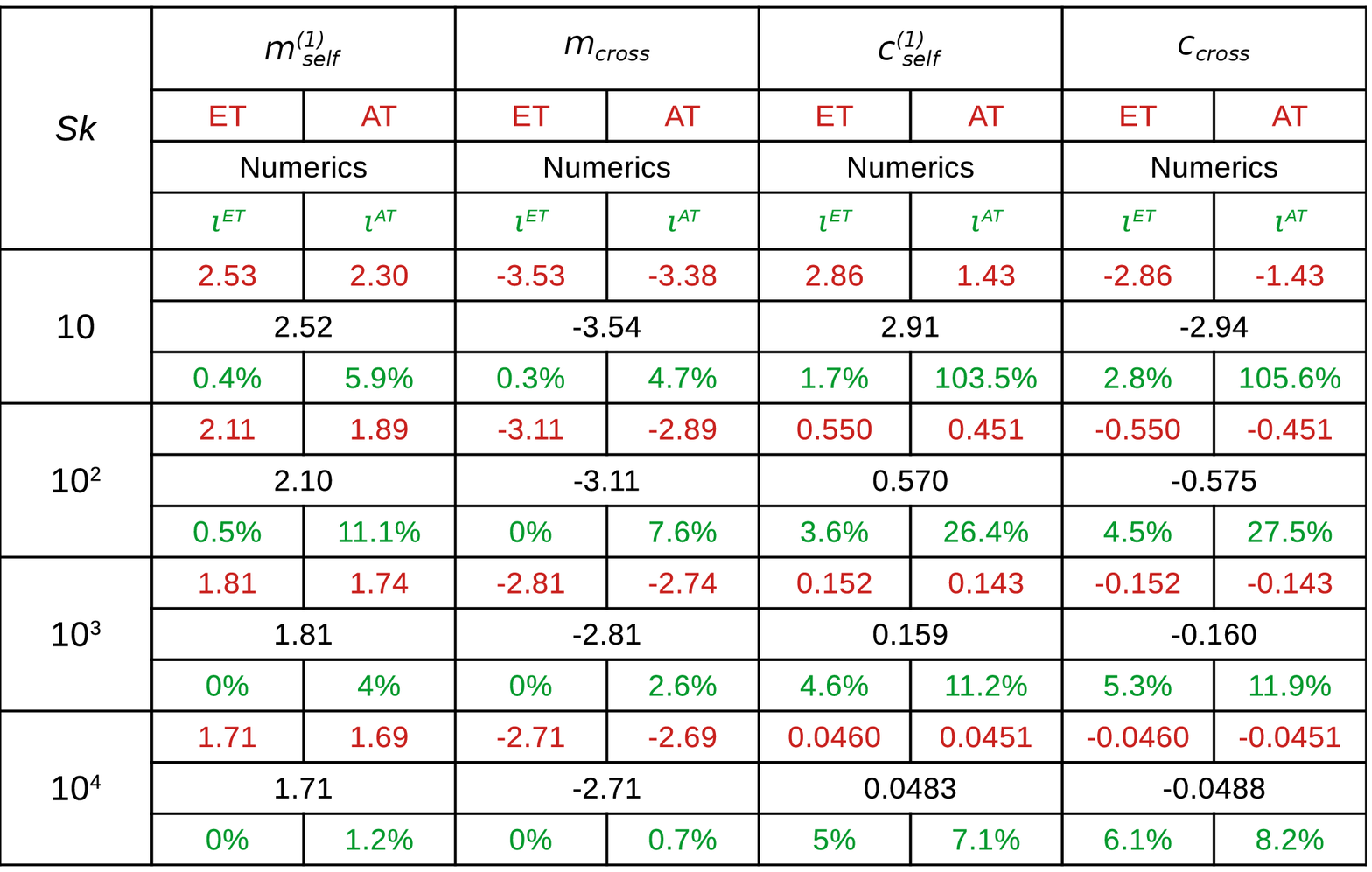}
    \caption{Study 1. Case 1. Table of the fluid added coefficients as a function of the Stokes number, $Sk$ and relative deviation, $\iota$. The notations ET and AT refer to the exact and asymptotic theories. The numerical values refer to symbols shown in Fig.~\ref{fig:coefficients_Sk}. The dimensionless separation distance is $\varepsilon=2$ and the Keulegan-Carpenter number is $KC = 10^{-2}$.}\label{tab:Comparison_theory_numerics_Sk}
\end{center}
\end{table}
%%%%%%%%%%%%%%%%%%%%%%%%%%%%%%%%%%%%%%%%%%%%%%%%%%%%%%%%%%
% Case 1. Evolution of added coefficients with Eps
%%%%%%%%%%%%%%%%%%%%%%%%%%%%%%%%%%%%%%%%%%%%%%%%%%%%%%%%%%
\begin{table}[!ht]
\begin{center}
    \includegraphics[width=0.675\linewidth]{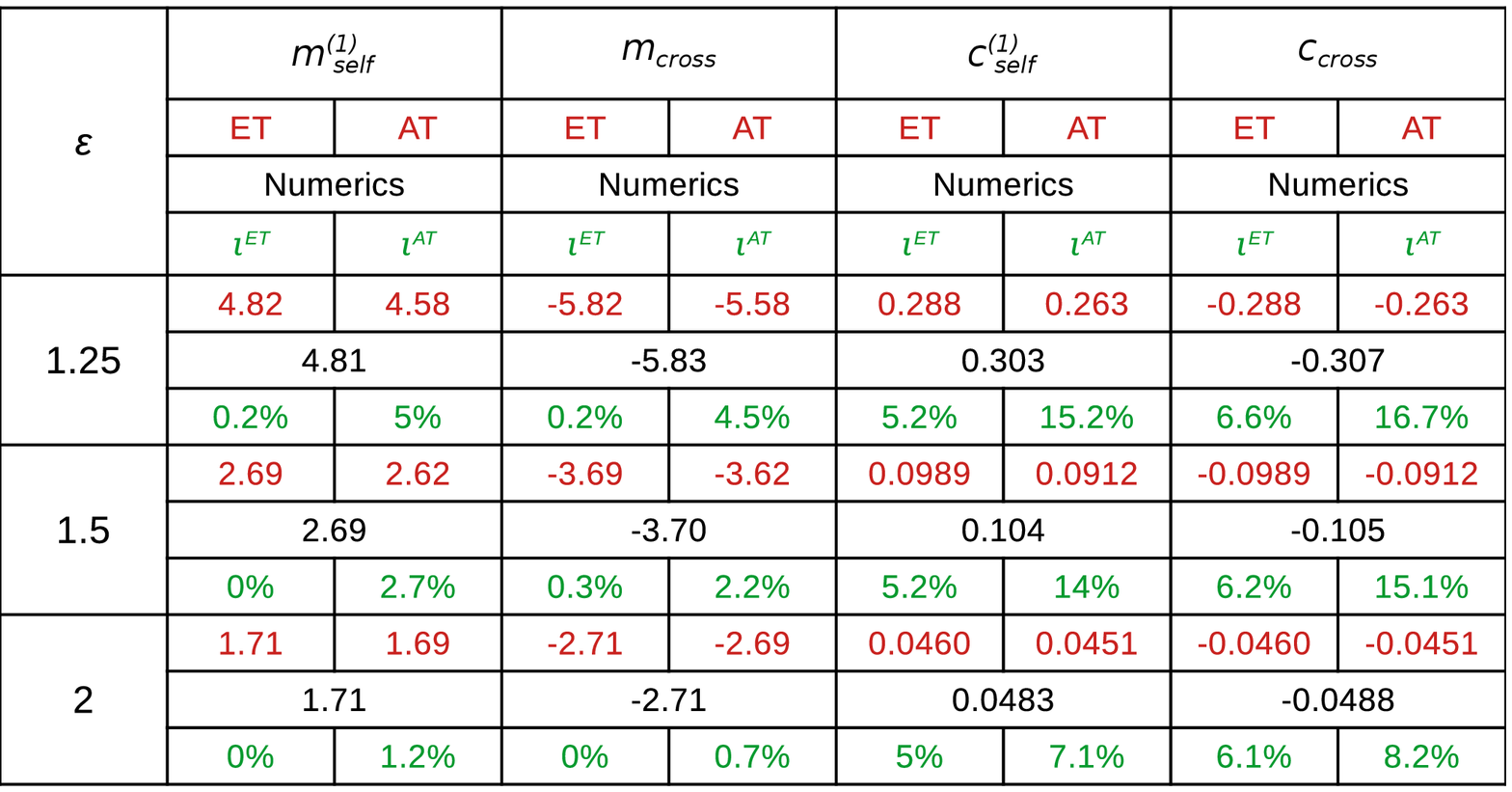}
    \caption{Study 1. Table of the fluid added coefficients as a function of the dimensionless separation distance, $\varepsilon$ and relative deviation, $\iota$. The notations ET and AT refer to the exact and asymptotic theories. The numerical values refer to symbols shown in Fig.~\ref{fig:coefficients_eps}. The Stokes number is $Sk=10^4$ and the Keulegan-Carpenter number is $KC = 10^{-2}$.}\label{tab:Comparison_theory_numerics_eps}
\end{center}
\end{table}
%%%%%%%%%%%%%%%%%%%%%%%%%%%%%%%%%%%%%%%%%%%%%%%%%%%%%%%%%%
% Case 1. Test of invariance
%%%%%%%%%%%%%%%%%%%%%%%%%%%%%%%%%%%%%%%%%%%%%%%%%%%%%%%%%%
\begin{table}[!ht]
\begin{center}
    \includegraphics[width=0.675\linewidth]{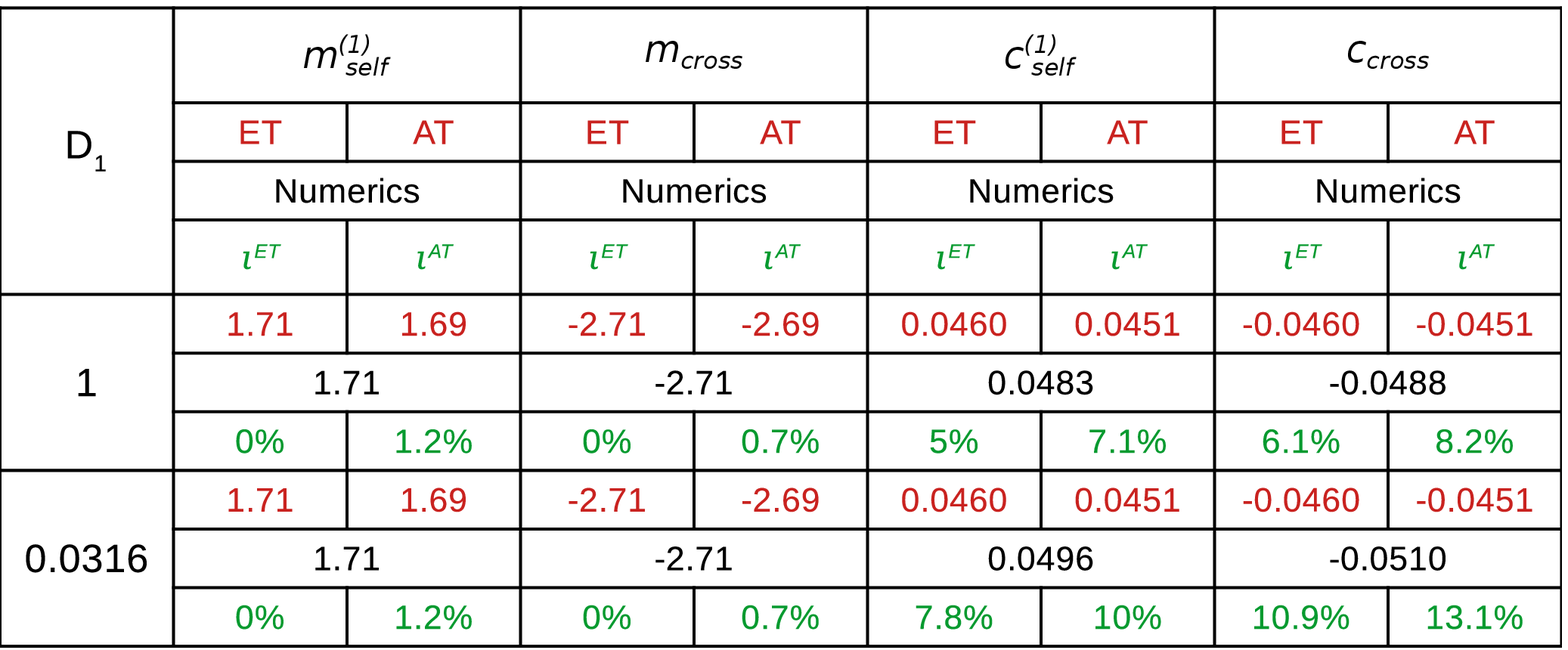}
    \caption{Study 1. Testing the scale invariance. Table of the fluid added coefficients and the relative deviation, $\iota$. The notations ET and AT refer to the exact and asymptotic theories. The numerical values are extracted from the dimensionless fluid forces shown in Fig.~\ref{fig:dimensionless_force_inv}. The Stokes number is $Sk=10^4$, the dimensionless separation distance is $\varepsilon=2$ and the Keulegan-Carpenter number is $KC = 10^{-2}$.}\label{tab:Comparison_theory_numerics_invariance}
\end{center}
\end{table}
%%%%%%%%%%%%%%%%%%%%%%%%%%%%%%%%%%%%%%%%%%%%%%%%%%%%%%%%%%
%Case 2. Evolution of cross forces: check symmetry of matrices
%%%%%%%%%%%%%%%%%%%%%%%%%%%%%%%%%%%%%%%%%%%%%%%%%%%%%%%%%%
\begin{table}[!ht]
\begin{center}
    \includegraphics[width=0.675\linewidth]{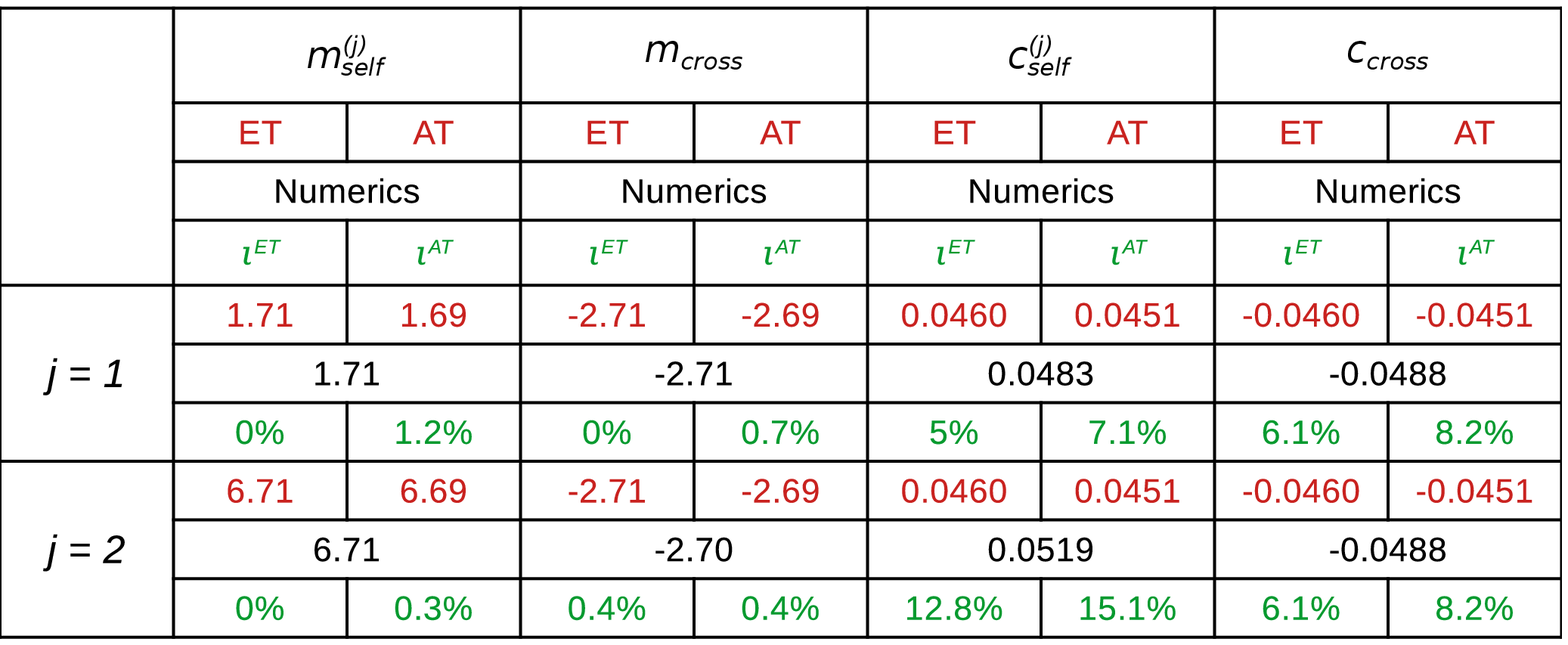}
    \caption{Study 1. Testing the symmetry of the fluid added mass and damping matrices. Table of the fluid added coefficients and the relative deviation, $\iota$. The notations ET and AT refer to the exact and asymptotic theories. The numerical values are extracted from the dimensionless fluid forces shown in Fig.~\ref{fig:dimensionless_force_outer}(b). The Stokes number is $Sk=10^4$, the dimensionless separation distance is $\varepsilon=2$ and the Keulegan-Carpenter number is $KC = 10^{-2}$.}\label{tab:Comparison_theory_numerics_case1_case2}
\end{center}
\end{table}

\section{Effect of the mesh size and time step on the fluid added coefficients} \label{sec:effect_mesh_size}
In this Appendix, we report the mesh converge analysis performed for both cases study. For the case study 1, in Tabs. \ref{tab:Mesh_properties_Sk_eps125_convergence}, \ref{tab:Mesh_properties_Sk_eps150_convergence}, \ref{tab:Mesh_properties_Sk_eps200_convergence}, \ref{tab:Mesh_properties_inv_convergence} and \ref{tab:Mesh_properties_outer_convergence} we clearly show a convergence of the mass coefficients as $l_c$ and $lc_{fine}$ are changed. The convergence of the damping coefficients is less obvious, especially for high values of $Sk$. Physically, this is related to the thickness of the boundary layer, which tends to zero as $Sk$ increases. It follows that a finer mesh is required close to a cylinder boundary to account for the thickness of the boundary layer and obtain an accurate estimation of the damping terms. Moreover, all the simulations are performed with $CFL = 1$ to ensure maximum precision, except for the simulations with a low value of the Stokes number, $Sk\in\{10^1,10^2\}$ and $\varepsilon = 2$ where the $CFL$ is progressively increased from 1 to 10, without influencing the quality of the results. Following the numerical approach of the case study 1, we perform a mesh sensitivity analysis on the case study 2, see Tab.~\ref{tab:Mesh_properties_Sk_convergence}.
%%%%%%%%%%%%%%%%%%%%%%%%%%%%%%%%%%%%%%%%%%%%%%%%%%%%%%%%%%%%%%%%%%%%%%%%%%%%%%%%%%%%%%%%%%%%%%%%%
\begin{table}[H]
\begin{center}
\includegraphics[width=0.675\linewidth]{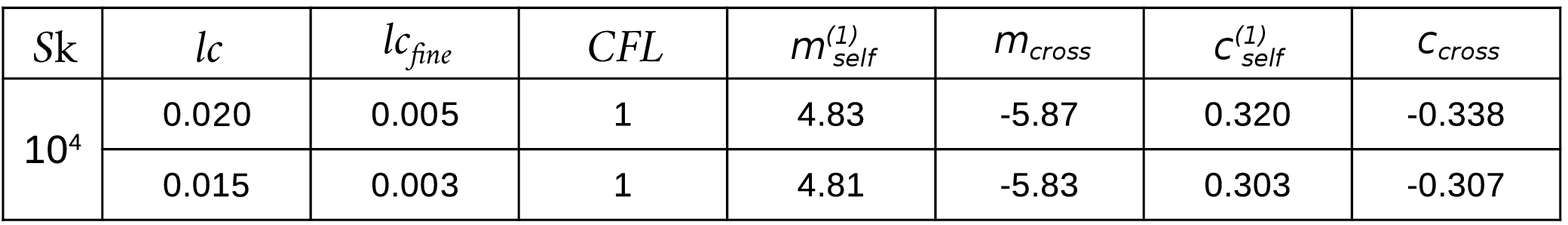}
    \caption{Study 1. Effect of the mesh size on the fluid added coefficients. The dimensionless separation distance is $\varepsilon=1.25$ and the Keulegan-Carpenter number is $KC = 10^{-2}$.}
    \label{tab:Mesh_properties_Sk_eps125_convergence}
\end{center}
\end{table}
%%%%%%%%%%%%%%%%%%%%%%%%%%%%%%%%%%%%%%%%%%%%%%%%%%%%%%%%%%%%%%%%%%%%%%%%%%%%%%%%%%%%%%%%%%%%%%%%%
\begin{table}[H]
\begin{center}
\includegraphics[width=0.675\linewidth]{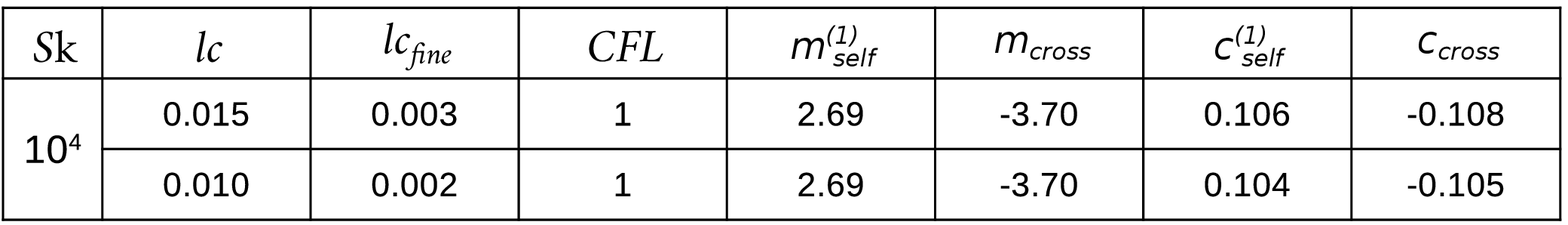}
    \caption{Study 1. Effect of the mesh size on the fluid added coefficients. The dimensionless separation distance is $\varepsilon=1.5$ and the Keulegan-Carpenter number is $KC = 10^{-2}$.}
    \label{tab:Mesh_properties_Sk_eps150_convergence}
\end{center}
\end{table}
%%%%%%%%%%%%%%%%%%%%%%%%%%%%%%%%%%%%%%%%%%%%%%%%%%%%%%%%%%%%%%%%%%%%%%%%%%%%%%%%%%%%%%%%%%%%%%%%%
\begin{table}[H]
\begin{center}
\includegraphics[width=0.675\linewidth]{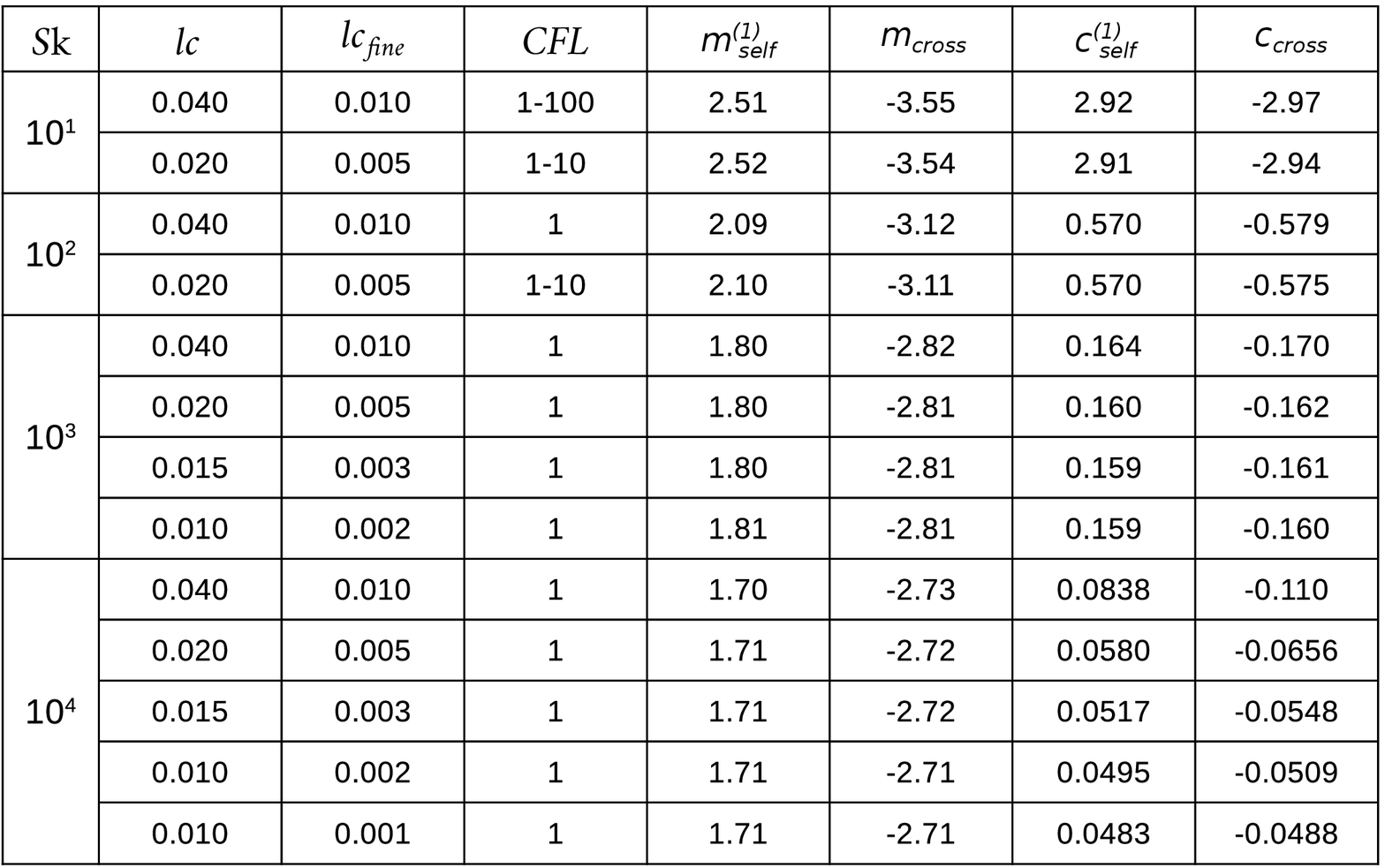}
    \caption{Study 1. Effect of the mesh size on the fluid added coefficients. The dimensionless separation distance is $\varepsilon=2$ and the Keulegan-Carpenter number is $KC = 10^{-2}$.}
    \label{tab:Mesh_properties_Sk_eps200_convergence}
\end{center}
\end{table}
%%%%%%%%%%%%%%%%%%%%%%%%%%%%%%%%%%%%%%%%%%%%%%%%%%%%%%%%%%%%%%%%%%%%%%%%%%%%%%%%%%%%%%%%%%%%%%%%%
\begin{table}[H]
\begin{center}
\includegraphics[width=0.675\linewidth]{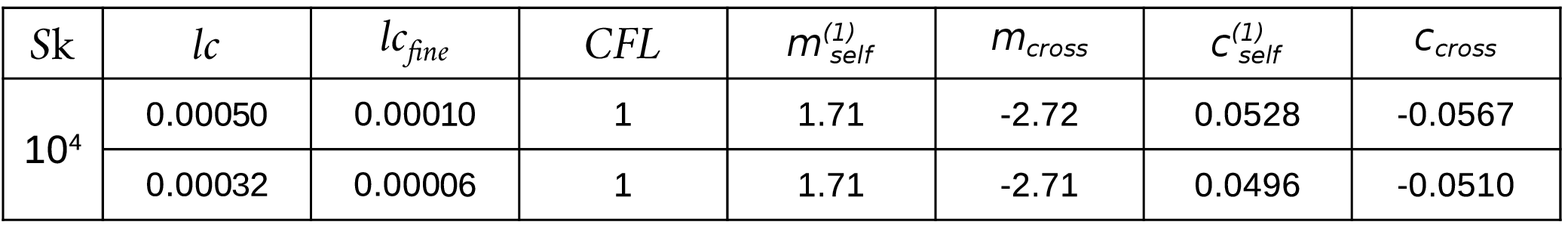}
    \caption{Study 1. Testing the scale invariance. Effect of the mesh size on the fluid added coefficients. The dimensionless separation distance is $\varepsilon=2$ and the Keulegan-Carpenter number is $KC = 10^{-2}$.}
    \label{tab:Mesh_properties_inv_convergence}
\end{center}
\label{tab:Meshproperties_inv}
\end{table}
%%%%%%%%%%%%%%%%%%%%%%%%%%%%%%%%%%%%%%%%%%%%%%%%%%%%%%%%%%%%%%%%%%%%%%%%%%%%%%%%%%%%%%%%%%%%%%%%%
\begin{table}[H]
\begin{center}
\includegraphics[width=0.675\linewidth]{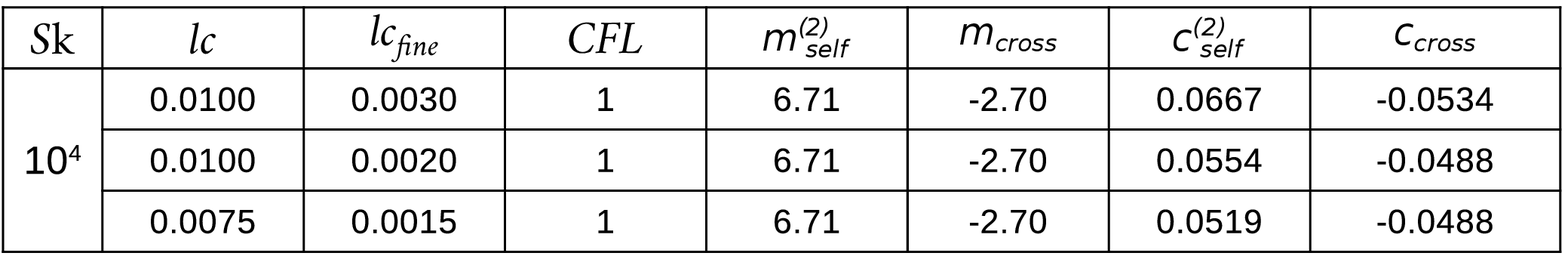}
    \caption{Study 1. Testing the symmetry of the fluid added mass and damping matrices. Effect of the mesh size on the fluid added coefficients. The dimensionless separation distance is $\varepsilon=2$ and the Keulegan-Carpenter number is $KC = 10^{-2}$.}
    \label{tab:Mesh_properties_outer_convergence}
\end{center}
\end{table}
%%%%%%%%%%%%%%%%%%%%%%%%%%%%%%%%%%%%%%%%%%%%%%%%%%%%%%%%%%%%%%%%%%%%%%%%%%%%%%%%%%%%%%%%%%%%%%%%%%%%
%%%%%%%%%%%%%%%%%%%%%%%%%%%%%%%%%%%%%%%%%%%%%%%%%%%%%%%%%%%%%%%%%%%%%%%%%%%%%%%%%%%%%%%%%%%%%%%%%
\begin{table}[H]
\begin{center}
\includegraphics[width=0.7\linewidth]{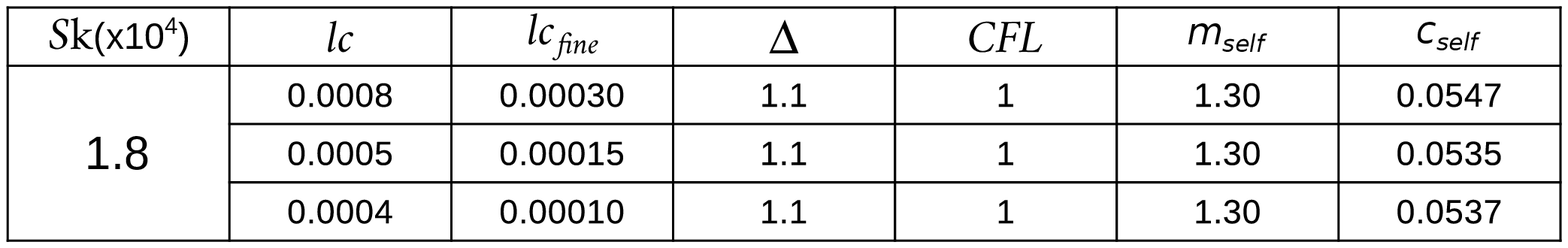}
    \caption{Study 2. Effect of the mesh size on the fluid added coefficients. The pitch ratio is $P/D=1.5$ and the Keulegan-Carpenter number is $KC = 10^{-1}$.}
    \label{tab:Mesh_properties_Sk_convergence}
\end{center}
\end{table}
%%%%%%%%%%%%%%%%%%%%%%%%%%%%%%%%%%%%%%%%%%%%%%%%%%%%%%%%%%%%%%%%%%%%%%%%%%%%%%%%%%%%%%%%%%%%%%%%%%%

%\bibliographystyle{model6-num-names}
\bibliographystyle{elsarticle-num}
\bibliography{biblio}

\end{document}